\documentclass{article}

\usepackage{microtype}
\usepackage{graphicx}
\usepackage{subcaption}
\usepackage{booktabs} % for professional tables
\usepackage{bm} % for bold math symbols
\usepackage[normalem]{ulem} % for strikethrough text
\usepackage{cancel}

\usepackage{hyperref}

\usepackage[preprint]{icml2026}

\usepackage{amsmath}
\usepackage{amssymb}
\usepackage{mathtools}
\usepackage{amsthm}
\usepackage{tikz-cd}
\usepackage{wrapfig}
\usepackage{enumerate}

\usepackage[capitalize,noabbrev]{cleveref}

\theoremstyle{plain}

\theoremstyle{definition}

\theoremstyle{remark}

\usepackage[textsize=tiny]{todonotes}

\icmltitlerunning{Weight-Space Physics: Interpretable Hypernetworks for Lattice Quantum Field Theories}

\begin{document}

\twocolumn[
  \icmltitle{Weight-Space Physics: \\Interpretable Hypernetworks for Lattice Quantum Field Theories}


  \icmlsetsymbol{equal}{*}

  \begin{icmlauthorlist}
    \icmlauthor{Tobias Göbel}{equal,uva-iop,ntu-phys}
    \icmlauthor{Julian R. Ebelt}{equal,uva-iop,ntu-phys}
    \icmlauthor{Zier Mensch}{equal,uva-iop,ntu-phys}
    \icmlauthor{Mathis Gerdes}{mit,iaifi}
    \icmlauthor{Miranda C.N. Cheng}{uva-iop,as-math}
  \end{icmlauthorlist}

  \icmlaffiliation{uva-iop}{Institute of Physics, University of Amsterdam, Amsterdam, Netherlands}
  \icmlaffiliation{as-math}{Institute for Mathematics, Academia Sinica, Taipei, Taiwan}
  \icmlaffiliation{ntu-phys}{Department of Physics, National Taiwan University, Taipei, Taiwan}
  \icmlaffiliation{mit}{Department of Physics, Massachusetts Institute of Technology, Cambridge, Massachusetts, USA}
  \icmlaffiliation{iaifi}{The NSF Institute for Artificial Intelligence and Fundamental Interactions}

  \icmlcorrespondingauthor{Miranda C.N. Cheng}{miranda.cheng.w@gmail.com}

  \icmlkeywords{
    Machine Learning, 
    ICML,
    Hypernetworks,
    Neural Network Interpretability,
    Representation Learning,
    Lattice Field Theory,
    Joint-Embedding Predictive Architecture,
    Normalizing Flows,
}

  \vskip 0.3in
]



\printAffiliationsAndNotice{\icmlEqualContribution}

\begin{abstract}
Lattice field theory is the workhorse of non-perturbative physics, used to simulate phenomena from the strong nuclear force to critical phenomena in materials. Its Boltzmann distributions are parametrized analytically by coupling constants, but these bare parameters are weak predictors of observables---extracting physics typically requires extensive simulation. While normalizing flows have emerged as effective samplers at fixed couplings, it remains difficult to interpret what these networks have learned. This raises a natural question: can the physics be read off directly from the flow \emph{network parameters themselves}, and can those parameters be generated for unseen theories? We propose lattice field theory as a testbed for neural network interpretability: because the target physics is qualitatively well-understood and smoothly varying, it provides ideal synthetic data with known ground truth. To this end, we introduce \textbf{JEPAWG}, a Joint-Embedding Predictive Architecture--based Weight Generator that maps couplings directly to flow weights via a learned latent space. On a scalar theory at lattices of size $6^2$ to $11^2$, the JEPAWG latent space recovers the correct intrinsic dimension of the underlying manifold, locates the phase transition, and encodes a finite-size shift aligned with the 2D Ising exponent $\nu \approx 1$, allowing us to uncover physical structure by studying the network weights alone. This suggests the fascinating idea of treating the network weights as a new type of physical observable. 
As a generator, JEPAWG also interpolates and extrapolates to unseen couplings effectively and remains robust to weight-space discontinuities introduced by multi-seed training data, outperforming PCA, AE, and VAE baselines.
\end{abstract}

\section{Introduction}

Lattice field theory (LFT) discretizes a quantum field onto a finite lattice, providing a first-principles approach to non-perturbative phenomena across high-energy and condensed matter physics. The action $S_g(\phi)$ is parametrized by bare coupling constants $g$, and physics is extracted by sampling the Boltzmann distribution $p_g(\phi) \propto e^{-S_g(\phi)}$ numerically.

Flow-based generative models have emerged as samplers for these
distributions \citep{albergo2021introduction,kanwar2024flow,cheng2026lecture}, learning an invertible map
from a tractable base to $p_g(\phi)$ and producing independent samples
that sidestep MCMC autocorrelation. A trained flow at coupling $g$ is
parametrized by a vector $\theta_g$, and we expect these parameters to carry
information about the physical structure of the theory. For instance, near a phase
transition, where $p_g$ varies rapidly with $g$, $\theta_g$ must respond
accordingly. This raises two complementary questions.
\emph{As a
diagnostic}---our primary focus--- does the geometry of these parameters in the weight space recover physical structure that the bare couplings themselves obscure?
\emph{As a
generator}, instrumentally, can $\theta_g$ be predicted from $g$, amortizing training
across couplings and extrapolating to unseen ones?

Lattice field theory is a uniquely well-suited testbed for these questions of \emph{weight interpretation} and \emph{weight generation}, because it supplies the kind of ground truth that interpretability research typically lacks. Understanding what neural networks have learned is a
central open problem \citep{rauker2023transparentaisurveyinterpreting, sharkey2025openproblemsmechanisticinterpretability}, but progress is often
bottlenecked by the absence of ground truth: for typical ML tasks, we
cannot independently verify what structure a learned representation
\emph{should} encode. Physical theories supply this ground truth. The target
distributions $p_g(\phi)$ are known analytically, the relevant coupling space
is low-dimensional and naturally smooth, and the renormalization
group makes quantitative predictions about its phase structure
(\Cref{subsec:cond-flow}). A representation of weight space can
therefore be checked against falsifiable physical predictions, not only
against qualitative intuitions. Physics, in this sense, acts as a supplier of natural synthetic data: rich enough to be non-trivial, structured enough to be interpretable.

To this end, we introduce \textbf{JEPAWG}, a JEPA-based Weight Generator
\citep{lecun2022path, assran2023ijepa} that maps couplings to flow parameters through a learned latent space. The same architecture that enables weight generation at unseen couplings also yields a latent representation whose geometry can be probed for physical structure. We apply JEPAWG to two-dimensional $\phi^4$ theory and make four contributions. 
\begin{enumerate}[(i)]
    \item \textbf{Recovery of physical dimensionality:} the JEPAWG latent space matches the 2D coupling manifold (MLE intrinsic dimension $2.01 \pm 0.06$; $98.5\%$ of variance in two principal axes), whereas AE and VAE baselines inflate this to $3.68$ and $4.88$. 
    \item \textbf{Phase transition from weights alone:} the pullback metric on $\theta(g)$ exhibits a ridge tracking the critical line of $\phi^4$ theory. 
    \item \textbf{Scale transformation from latent geometry:} comparing weights of flows trained at different lattice sizes through the learned embedding, and requiring physically similar theories to lie nearby in the embedding space, we find that the latent geometry is organized by the finite-size scaling variable, with the optimal exponent broadly consistent with the 2D Ising value~$\nu=1$. 
    \item \textbf{Generation:} JEPAWG produces functional flows at unseen couplings, reaching an effective sample size (ESS) of $0.88$ on multi-seed interpolation, where PCA, AE, and VAE baselines reach only $0.30$--$0.40$, and it remains robust under the weight canonicalization procedure.

\end{enumerate}

\section{Related Work}
\paragraph{Joint-embedding predictive architectures.}
Joint-embedding predictive architectures (JEPA) learn representations by predicting in latent space rather than reconstructing the input~\citep{lecun2022path}. I-JEPA~\citep{assran2023ijepa} and V-JEPA~\citep{bardes2024revisiting} demonstrated this for images and video, while VL-JEPA~\citep{chen2025vljepa} extended the framework to vision-language tasks, attaching a lightweight decoder to translate predicted embeddings back into text. To our knowledge, JEPA has not previously been applied to neural network weight spaces. 

\paragraph{Hypernetworks and weight generation.}
A hypernetwork~\citep{schmidhuber1992learning, ha2017hypernetworks} is a network that produces the weights of another, conditioned on a task or system descriptor. This paradigm has been applied to continual learning~\citep{vonOswald2020continual} and approximate Bayesian inference over weights~\citep{krueger2017bayesian}; see \citet{chauhan2024hypernetworks} for a survey. Our work instantiates this paradigm for flow models simulating quantum field theories: a learned encoder--predictor--decoder pipeline outputs per-coupling weights, enabling extrapolation to unseen couplings without retraining.

\paragraph{Flow-based sampling for lattice field theory.}
Normalizing flows have been widely applied as samplers for lattice field theories; see e.g.~\citet{cheng2026lecture,kanwar2024flow} for reviews. Closest to our setting, \citet{gerdes2023learning} train a single coupling-conditional flow that generalizes across couplings by conditioning each layer on $g$ while keeping network weights shared. Instead of conditioning a fixed network, our method treats trained flow parameters as data and learns a latent embedding that decodes to flow parameters at unseen couplings.

\section{Background}

\subsection{Scalar Quantum Field Theory on the Lattice}
\label{subsec:cond-flow}
We work with the two-dimensional real scalar $\phi^{4}$ theory on a
periodic square lattice $V_{L}\cong(\mathbb{Z}/L\mathbb{Z})^{2}$, with
site variables $\phi_{x}\in\mathbb{R}$ and action
\begin{equation}
    S_{g}(\phi)
    = \sum_{ x,y}\phi_{x}\Delta_{xy}\phi_{y}
    + \sum_{x\in V_{L}} m^{2}\phi_{x}^{2}
    + \sum_{x\in V_{L}} \lambda\phi_{x}^{4},
    \label{eq:phi4action}
\end{equation}
where $\Delta$ is the discrete Laplacian and
$g=(m^{2},\lambda)$ collects the two bare couplings\footnote{These are called \emph{bare} couplings because they appear directly in the action, before quantum effects are taken into account; the physically meaningful indicators are the observables whose mapping to bare couplings is typically very complex and 
depends on the lattice spacing.}. Physical
observables are expectations under the Boltzmann measure
$p_{g}(\phi)\propto e^{-S_{g}(\phi)}$. Despite its simplicity,
this theory exhibits a non-trivial phase structure and therefore forms an excellent testbed for studying what neural samplers learn, since the known phase structure provides a falsifiable ground truth against which learned representations can be checked.

\paragraph{Phase structure.}
Near a phase transition, the physics becomes acutely sensitive to the couplings: a small change in $g$ produces a qualitative change in the theory, which makes critical regions among the hardest to simulate. At the same time, the long-distance physics there becomes \emph{universal}: it is determined only by symmetry and dimensionality, independent of microscopic details. Together, these properties make critical points a natural place to test whether a learned representation has captured the physical structure.

The action~\eqref{eq:phi4action} is invariant under the global
$\mathbb{Z}_{2}$ symmetry $\phi\mapsto-\phi$. For $\lambda>0$,
the theory has two phases separated by a critical line
$m^{2}=m^{2}_{c}(\lambda)$.
When $m^2 > m^2_c(\lambda)$, the lattice magnetization $\bar\phi=L^{-2}\sum_{x}\phi_{x}$ averages
to zero (the \emph{paramagnetic} phase).
When $m^{2}$ is sufficiently negative, the potential acquires
two degenerate minima; the system settles into one of them, breaking
$\mathbb{Z}_{2}$ spontaneously, and $\langle|\bar\phi|\rangle\neq 0$
(the \emph{ferromagnetic} phase). The coupling space $(m^2,\lambda)$
is two-dimensional and the two phases are separated by a
one-dimensional critical line; any representation that captures the
physics should reflect this low-dimensional
structure.

\paragraph{Observables.}
The transition is conventionally diagnosed by moments of the
magnetization. The susceptibility
$\chi(g)=L^{2}\bigl(\langle\bar\phi^{2}\rangle-\langle|\bar\phi|\rangle^{2}\bigr)$
peaks along the critical line, while the Binder cumulant
$U_{4}(g)=1-\langle\bar\phi^{4}\rangle/(3\langle\bar\phi^{2}\rangle^{2})$
interpolates between $0$ in the paramagnetic phase and $2/3$ in the
ferromagnetic one, with curves at different $L$ crossing at the
critical point. Two-dimensional $\phi^4$ theory belongs to the 2D
Ising universality class, with the correlation-length critical exponent $\nu = 1$.
A learned representation that has captured the physical content of the
theory should therefore organize the coupling plane by RG-invariant
combinations rather than by $g$ itself, a property we test in
\Cref{sec:interpretability}.

\paragraph{Normalizing flows for lattice field theory.}
Estimating these observables requires drawing samples from the Boltzmann measure $p_g(\phi)\propto e^{-S_g(\phi)}$, a task that becomes increasingly difficult near the critical line, where local Monte Carlo updates suffer from critical slowing down. In recent years, flow-based methods have been introduced as successful samplers of lattice field theories \citep{albergo2019flow}, offering the promise of mitigating critical slowing down.
A normalizing flow~\citep{tabak2010DensityEstimation,papamakarios2021NormalizingFlows, kobyzev2020normalizing} learns an invertible map $f_\theta : \mathbb{R}^{L^2} \to \mathbb{R}^{L^2}$ that transforms a tractable base distribution $\rho$ (typically a standard Gaussian) into an approximation of the target Boltzmann measure $p_g(\phi)$. The map is trained by minimizing the reverse KL divergence between the pushforward of $\rho$ under $f_\theta$ and $p_g$, using only the known action $S_g$ and samples from the flow itself. A conditional flow architecture~\citep{gerdes2023learning} extends this by making the parameters depend smoothly on the couplings $g = (m^2, \lambda)$, so that a single model produces a sampler for each theory in a continuous region of coupling space.

\subsection{Joint-embedding predictive architectures}
\label{subsec:jepa}
A joint-embedding predictive architecture (JEPA)
\citep{assran2023ijepa, lecun2022path} learns representations by predicting the
embedding of one signal from the embedding of a related one, rather
than reconstructing the input directly. Given two such signals, $x$ and $y$, that describe the same underlying object,
encoders $f_x$ and $f_y$ produce embeddings $z_x = f_x(x)$,
$z_y = f_y(y)$ and a predictor $h$ maps between them. The encoders
and the predictor are trained jointly, so that the embeddings and
the predictive map between them are shaped together by a single
objective enforcing $h(z_y) \approx z_x$.

Unlike generative architectures, JEPA operates
entirely in representation space, encouraging the embeddings to capture
high-level structure rather than input-level detail. In our setting
(\Cref{sec:jepawg}), the underlying object is a lattice theory at a fixed coupling, and the two signals are its bare coupling $g$ and a flow's parameter vector $\theta$ trained to sample at $g$.

Without additional regularization, both encoders can collapse to a
constant mapping, trivially satisfying the alignment objective. The
VICReg objective \citep{bardes2022vicreg} prevents this with three terms:
an \emph{invariance} term
$s = \tfrac{1}{N}\|h(z_y) - z_x\|_2^2$ that enforces alignment, a
\emph{variance} term that penalizes small variance and hence representational collapse,
and a
\emph{covariance} term that penalizes correlations between latent
dimensions and thereby encourages the utilization of all available latent dimensions.
Full expressions are given in
Appendix~\ref{app:vicreg}.

\section{Methods}
We introduce JEPAWG (\Cref{sec:jepawg}), our JEPA-based weight
generator that maps couplings to flow parameters through a learned
latent space, and compare it against three baselines (PCA, AE, VAE), trained without coupling information (\Cref{app:baseline_methods}).
\subsection{JEPAWG}
\label{sec:jepawg}
We slightly modify the joint-embedding predictive architecture (JEPA) introduced
in \Cref{subsec:jepa}, and
augment it with a decoder that enables \emph{weight
generation} at unseen couplings. We refer to the resulting model as
JEPAWG (JEPA-based Weight Generator), combining JEPA-style
representation learning with a hypernetwork-style mapping from coupling
space to flow parameters.
We use a JEPA backbone, rather than a coupling-conditional regressor
$g \mapsto \hat{\theta}$ trained with reconstruction loss, because JEPA
shapes the latent space from both signals jointly: the predictor is
trained to recover the weight embedding $\mathbf{z}_\theta$ from the
coupling embedding $\mathbf{z}_g$, while the variance term prevents
either embedding from collapsing to a constant.
The latent space is therefore incentivized to align with the coupling structure during training rather than fitted to it
post-hoc. We will see that this feature is most helpful when the relevant sub-manifold in the weight space is non-smooth (\Cref{sec:weight_generation}).

\begin{figure}
\centering
\begin{tikzcd}[column sep=1.5cm, row sep=1.2cm]
  \theta \arrow[d, bend left, "E_\theta"]
  & g \arrow[d, "E_g"] \\
  \mathbf{z}_\theta \arrow[u, dashed, bend left, "D_\theta"]
  & \mathbf{z}_g \arrow[l, "P"']
\end{tikzcd}
\caption{JEPAWG architecture: weight encoder $E_\theta$, coupling encoder $E_g$, predictor $P$, and decoder $D_\theta$.}
\label{fig:jepawg-arch}
\end{figure}
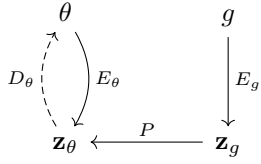

\paragraph{Architecture.}
The model consists of four components: (i) a weight encoder
$E_\theta : \mathbb{R}^{d_\theta} \to \mathbb{R}^{d_z}$ (with $d_\theta = 19{,}689$, the number of neural ODE weights; see Appendix~\ref{app:dg:cond}) mapping canonicalized\footnote{See \Cref{sec:data_prep}} 
parameter vectors $\theta$ to weight embeddings $\mathbf{z}_\theta = E_\theta(\theta)$,
(ii) a coupling encoder $E_g : \mathbb{R}^2 \to \mathbb{R}^{d_z}$ mapping
couplings $g = (m^2,\lambda)$ to coupling embeddings $\mathbf{z}_g = E_g(g)$,
(iii) a predictor $P : \mathbb{R}^{d_z} \to \mathbb{R}^{d_z}$ mapping
coupling embeddings into the weight embedding space, trained so that
$P(\mathbf{z}_g) \approx \mathbf{z}_\theta$, and (iv) a decoder
$D_\theta : \mathbb{R}^{d_z} \to \mathbb{R}^{d_\theta}$ mapping predicted weight
embeddings back to parameter vectors. At inference, the composition
$D_\theta \circ P \circ E_g : g \mapsto \hat{\theta}$ defines a hypernetwork
from couplings to flow parameters.

\paragraph{Training.}
The encoders $E_\theta, E_g$ and predictor $P$ are trained using the
VICReg objective~\eqref{eq:vicreg}, with the covariance term removed. The reason is that this term deliberately decorrelates latent dimensions and encourages the representation to spread across all $d_z$ available directions; however, we wish to retain and exploit the low-dimensional structure of the problem---our object of study is a two-parameter family of Boltzmann distributions. This property is probed directly in our experiments, as discussed 
in \Cref{sec:interpretability}. For completeness, we also evaluate the variant that retains the covariance term (i.e., the full VICReg objective), which we denote JEPAWG$^*$.
After training the encoders and the predictor,
we freeze $(E_\theta, E_g, P)$ and train the decoder $D_\theta$. This two-stage procedure ensures that the latent space
is shaped entirely by the JEPA objective; training the decoder jointly
would allow reconstruction pressure to leak into the weight encoder,
encouraging it to retain $\theta$-specific detail beyond what is
predictable from $g$. The decoder minimizes 
\begin{equation}
    \mathcal{L}_{\mathrm{dec}}
  = \frac{1}{N} \sum_{i=1}^{N}
    \bigl\| D_\theta(P(E_g(g_i))) - \theta_i \bigr\|_2^2
\end{equation}
on training pairs $\{(\theta_i, g_i)\}_{i=1}^N$. Throughout, we set
$d_z = 16$, deliberately over-sizing the latent space relative
to the two-dimensional coupling manifold so that any concentration
onto fewer dimensions is a learned property rather than an
architectural constraint.

\paragraph{Inference.}
Given a new coupling $g^\star$, we generate flow parameters by applying the map
 $\hat{\theta}(g^\star) = D_\theta\bigl(P(E_g(g^\star))\bigr)$.
The generated parameters $\hat{\theta}(g^\star)$ can then be used to instantiate a flow model at the desired coupling.

\subsection{Baselines}

We compare against three baselines: weight-space PCA, an autoencoder (AE) and a variational autoencoder (VAE); see \Cref{app:baseline_methods}. Each method operates on flattened parameter vectors $\theta \in \mathbb{R}^{d_\theta}$, standardized per dimension as $\tilde{\theta}_i = (\theta_i - \bar{\theta}_i)/\sqrt{\mathrm{Var}(\theta_i)}$. Coupling information enters only at inference, through a post-hoc regressor mapping $g = (m^2, \lambda)$ to the learned embedding, which allows the baseline to act as a conditional generator. All baselines are trained on the same parameter vectors as JEPAWG, isolating the architectural comparison from data-pipeline effects. We report results in \Cref{tab:split_ess} and discuss the asymmetric effect of canonicalization in \Cref{sec:weight_generation}. As an ablation, we also include JEPAWG$^*$, which uses the full VICReg loss with a nonzero covariance term.

\section{Experiments}
\label{sec:experiments}

A central question we address is whether a single framework can achieve both competent flow-model generation for unseen couplings and meaningful representation learning whose geometry reflects the underlying physics. We
evaluate JEPAWG on both criteria.
We focus our analysis on the multi-seed setting (see \Cref{app:data-generation})
where the training data
occupy disjoint regions of weight space, to ensure that the representations JEPAWG learns are not merely a result of the specific way of data generation.

For \textbf{weight generation}, we measure the effective sample size
(ESS;~\eqref{eqn:ess}) of decoded flows on interpolation and extrapolation points
(\Cref{sec:weight_generation}).
For \textbf{representation geometry},  we estimate the intrinsic
dimension of the latent space, visualize its structure, and show
that the geometry of the parameter map recovers the location of the
phase boundary directly, attributing the criticality signal to non-local components of
the network (\Cref{sec:phasetransition}).

\subsection{Data Preparation}
\label{sec:data_prep}
We use the conditional flow of \citet{gerdes2023learning} as the first step in our data-generation machinery. On a $19 \times 19$ grid in the space of couplings $(m^2,\lambda)$, 
we fine-tune the models generated by the conditional flow to have high ESS. We repeat the process for lattices with sizes $6^2$ to $11^2$ as described in \Cref{app:data-generation}. 
The network weights have various symmetries, including changes of basis, permutations, and sign flips, which change the
parameter vector $\theta$ while leading to the same input-output map.
We fix this
redundancy with a {\bf weight canonicalization} procedure: we map each $\theta$ to a unique canonical representative
per symmetry orbit (Appendix~\ref{app:canon}) to facilitate non-ambiguous feature extraction from weights; the importance of this step
is quantified in \Cref{sec:weight_generation}. 
More details of the data preparation procedure are
described in Appendix~\ref{app:data-generation}. 

\subsection{Weight Generation}
\label{sec:weight_generation}
In the data preparation stage,
we first train two conditional flows with independent random
initializations, each covering the full $\lambda$ interval but
a disjoint half of the $m^2$ range: seed 1 covers $m^2 \in [-5.1, -3.5]$ and seed 2 covers $m^2 \in (-3.5, -1.9]$, both over $\lambda \in [2.85, 5.65]$.
The training set for all
methods is assembled by combining the $(g, \theta_g)$ pairs  from
the two parts. Because the two conditional flows converge to different optima, the
weight manifold has a structural discontinuity at the boundary, even
though the underlying physics varies continuously. This multi-seed approach thus helps to ensure the robustness of our experiments: JEPAWG does not simply inherit coherent structure from the conditional network. 
We evaluate on
256 coupling pairs $(m^2,\lambda)$ uniformly sampled within their respective full training intervals, and 32 coupling pairs in each of eight extrapolation regions, each extending a distance $\delta = 1.5$ beyond the training window along the relevant axis (\Cref{fig:box_explanation}).
 These eight regions are divided into four edge regions (one of the coupling constants beyond the
training window) and four corner regions (both couplings outside the
training window); see \Cref{fig:box_explanation} for an illustration.
\Cref{tab:split_ess} shows that JEPAWG produces the strongest flows
in this setting, achieving an interpolation ESS of ${\bf 0.88}$.
JEPAWG$^*$ reaches $0.61$, while the reconstruction-trained baselines (VAE, AE,
PCA) lag substantially at $0.30$--$0.40$. The gap widens on
extrapolation: JEPAWG retains $0.56$ on edges and $0.39$ on corners,
where the weakest baselines fall below~$0.10$. The embeddings
(\Cref{fig:jepawg-vs-ae-latent}) offer insights into the structural reasons behind this result:
the AE and VAE encoders separate the two seeds into disjoint clusters
rather than forming a continuous manifold, while JEPAWG maps both
seeds onto a single smooth sheet. 

\begin{table}
\centering
\caption{Decoded-flow ESS on the split-seed grid for $L=6$. The coupling plane
is split into a training/interpolation window, four edge regions, and
four corner regions, as illustrated in \Cref{fig:box_explanation}.
Interpolation: ESS estimation on 256 uniformly sampled couplings
$(m^2,\lambda)$ with $m^2\in[-5.1,-1.9]$ and $\lambda\in[2.85,5.65]$.
Edges: mean$\pm$std over the four edge regions at $\delta = 1.5$
(32 points/region). Corners: mean$\pm$std over the four corner
regions at $\delta = 1.5$ (32 points/region). The first three columns
use canonicalized weights; ``Raw Interp.'' reports interpolation
ESS without weight canonicalization. See \Cref{tab:ess_split_seed} for the per-region breakdown.}
\label{tab:split_ess}
\small
\setlength{\tabcolsep}{3pt}
\begin{tabular}{lcccc}
\toprule
Method & Can. Interp. & Can. Edges & Can. Corner & Raw Interp. \\
\midrule
JEPAWG & $\mathbf{0.88{\pm}0.18}$ & $\mathbf{0.56{\pm}0.19}$ & $\mathbf{0.39{\pm}0.19}$ & $0.78{\pm}0.22$ \\
JEPAWG$^*$   & $0.61{\pm}0.32$ & $0.35{\pm}0.12$ & $0.25{\pm}0.09$ & $0.41{\pm}0.33$ \\
VAE    & $0.40{\pm}0.34$ & $0.23{\pm}0.07$ & $0.13{\pm}0.14$ & $0.91{\pm}0.18$ \\
AE     & $0.33{\pm}0.33$ & $0.15{\pm}0.05$ & $0.08{\pm}0.07$ & $\mathbf{0.92{\pm}0.18}$ \\
PCA    & $0.30{\pm}0.31$ & $0.09{\pm}0.02$ & $0.10{\pm}0.07$ & $0.75{\pm}0.27$ \\
\bottomrule
\end{tabular}
\end{table}

Recall that canonicalization selects a single, specific representative from
each orbit of symmetry-equivalent weights (Section~\ref{sec:data_prep},
Appendix~\ref{app:canon}). It is therefore the step that lets us
properly account for the weight-space symmetries before attempting to
interpret the weights. This comes at a price: collapsing each orbit onto its
canonical representative makes the weight map $\theta(g)$ vary less smoothly
with the couplings, sharpening its discontinuity across coupling space.
Table~\ref{tab:split_ess} reveals a dichotomy in how the methods respond to canonicalization. Not surprisingly, it hurts all of the non-JEPA methods,
which do not cope well with the induced discontinuity: raw-weight VAE, AE, and
PCA reach interpolation ESS of $0.910$, $0.915$, and $0.749$, but fall to
$0.40$, $0.33$, and $0.30$ once canonicalized. For JEPAWG, by contrast, canonicalization can help (JEPAWG: $0.78 \to 0.88$; JEPAWG$^*$:
$0.41 \to 0.61$): its latent space is shaped by the coupling signal rather
than by reconstructing the raw weights, so it is not degraded by fixing the
symmetry representative.
More importantly, this ESS comparison is not the whole story. In both
settings, raw and canonicalized weights, only JEPAWG produces an
interpretable latent space (intrinsic dimension matching the coupling
manifold, smooth extrapolation, and recovery of the critical exponent), whereas
the baselines do not, even at their best raw-weight setting. We discuss this
in the next subsection. A single-seed variant of this experiment is reported
in Appendix~\ref{sec:single_seed}.
\subsection{Interpretability}
\label{sec:interpretability}

\begin{table}
\caption{Intrinsic dimension estimates of $z_\theta$ at encoder width $H{=}128$ and
latent dimension $d_z{=}16$. Top-5 PCA denoising; mean $\pm$ standard
deviation across 5 seeds.}
\label{tab:intrinsic_dim}
\centering
\small
\setlength{\tabcolsep}{4pt}
\begin{tabular}{l cc}
\toprule
        & MLE (K=10)              & PCA-95\% \\
\midrule
dim$((m^2,\lambda))$     & $2.00$                  & $2$ \\
\midrule
JEPAWG  & $\mathbf{2.01{\pm}0.06}$ & $\mathbf{2.0{\pm}0.0}$ \\
JEPAWG$^*$    & $2.36{\pm}0.04$         & $13.2{\pm}0.8$ \\
VAE     & $4.88{\pm}0.15$         & $14.8{\pm}0.4$ \\
AE      & $3.68{\pm}0.09$         & $11.6{\pm}0.5$ \\
PCA     & $2.34{\pm}0.01$         & 15 \\
\bottomrule
\end{tabular}
\end{table}

\paragraph{Intrinsic dimension.}
We estimate the intrinsic dimension of each learned representation
using the maximum likelihood estimator (MLE) of
\citet{levina2004maximum} (details in
Appendix~\ref{app:analysis-tools}). \Cref{tab:intrinsic_dim} shows
that JEPAWG is the only method whose MLE estimate matches the
ground truth of $d{=}2$ (the dimension of the coupling space we used), and the only one whose latent variance is
captured almost entirely by two principal axes (PCA-95\%${=}2$). The
VAE and AE inflate the intrinsic dimension well beyond the physical
ground truth, with MLE estimates of $4.88$ and $3.68$ respectively,
indicating that their latent spaces retain structure unrelated to the
coupling parameters.

\begin{figure}
\centering
\includegraphics[width=\linewidth]{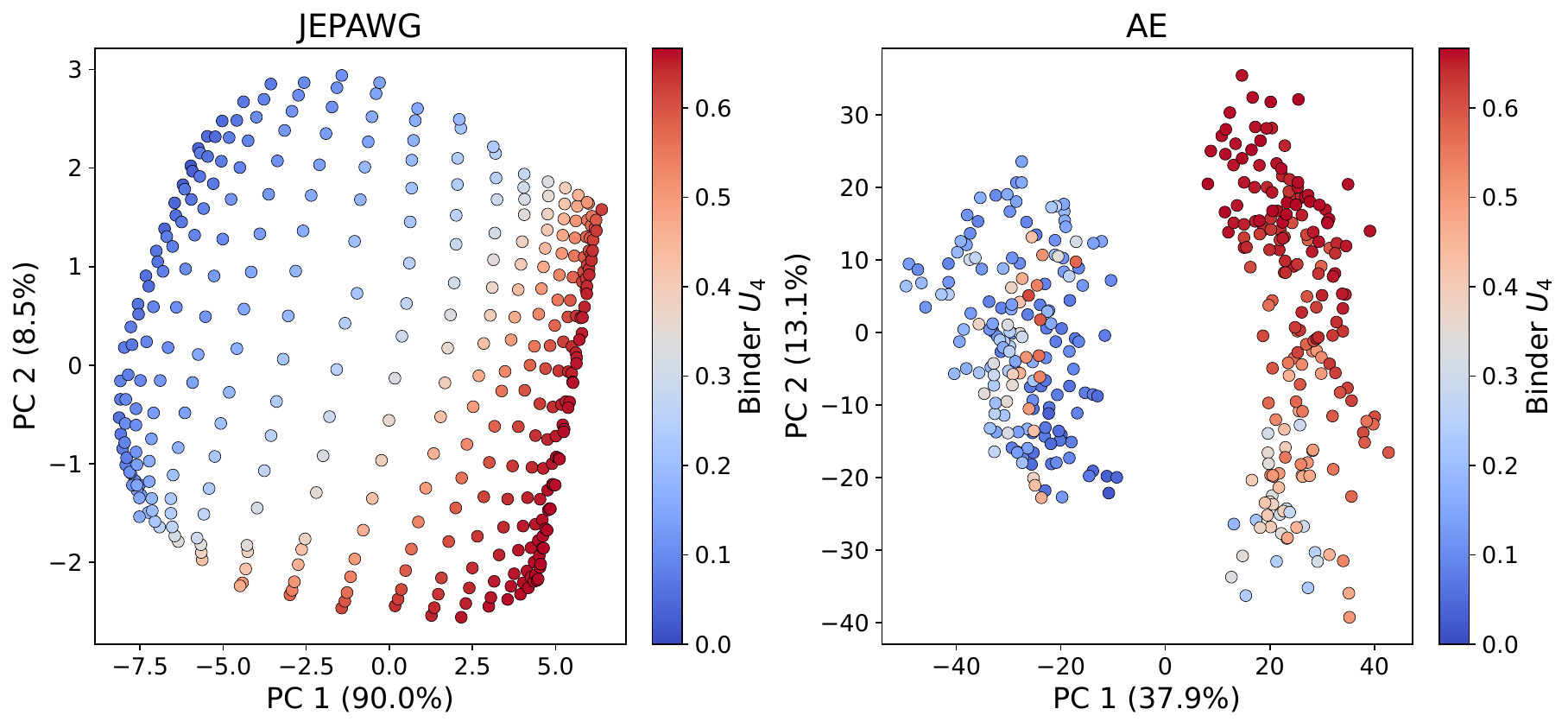}
\caption{PCA of latent representations ($d_z{=}16$) for JEPAWG (left) and
autoencoder (right), trained on 361 canonicalized flow grid points from
two independently trained source flows. Points are colored by the Binder
cumulant $U_4$. JEPAWG organizes the latent space along an axis aligned
with the phase structure, whereas the autoencoder embedding exhibits less
structure and splits into two disconnected components reflecting the
seed boundary inherited from the 2-seed setup with which the training data are generated.}

\label{fig:jepawg-vs-ae-latent}
\end{figure}

\paragraph{Latent-space embedding.}
\Cref{fig:jepawg-vs-ae-latent} shows the first two principal
components of each representation, colored by the Binder cumulant
$U_4$; see \Cref{fig:all-methods-latent} for the full comparison
across all five methods and additional colorings by $m^2$ and
$\lambda$. For JEPAWG, PC$_1$ and PC$_2$ capture
$98.5\%$ of the latent variance and form a smooth sheet aligned with
the coupling parameters; $U_4$, which is never seen during training,
varies coherently along the same axes. The autoencoder places only
$51.0\%$ of variance in its leading two components, and neither
coupling nor $U_4$ traces a coherent direction; instead, the two
training seeds separate into distinct clusters, indicating that the
encoder retains seed identity rather than learning a continuous
representation of the physics. This concentration of the JEPAWG
representation onto two physically relevant axes is consistent with
the PCA-$95\%{=}2$ result in \Cref{tab:intrinsic_dim}. The smoothness of the JEPAWG embedding also holds in the single-seed setting (\Cref{sec:single_seed}). 

\paragraph{Phase transition.}
\phantomsection
\label{sec:phasetransition}
We study how the parameter map $g \mapsto \theta(g)$
varies across coupling space by computing the pullback metric
$G_{ij}(g) = \langle \partial_i \theta(g), \partial_j \theta(g) \rangle$,
with $\langle\cdot,\cdot\rangle$ the Euclidean inner product on
the weight space and derivatives estimated by finite
differences in $(m^2, \lambda)$. The area element
$A(g) = \sqrt{\det G(g)}$ quantifies how fast the learned parameters
deform with the bare couplings; a peak is expected near the critical
line, where $\theta(g)$ must accommodate rapid changes in the physics.

\begin{figure*}
\centering
\includegraphics[width=0.9\linewidth]{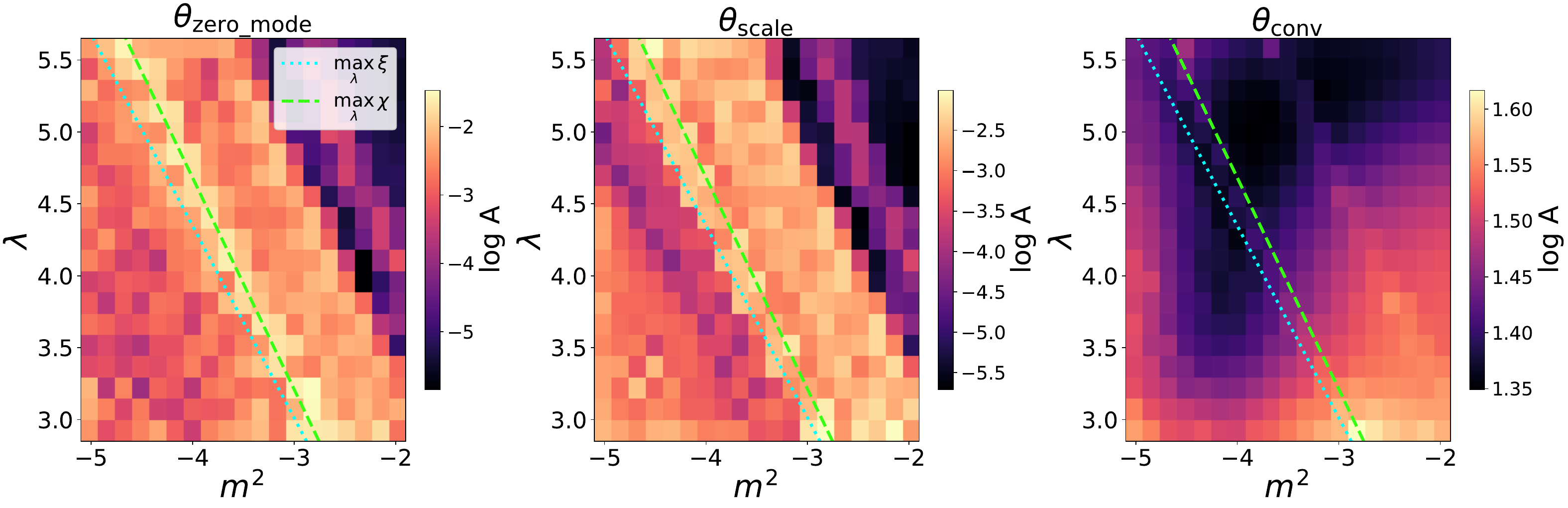}
\caption{Pullback area $A(g)$ on the $(m^2, \lambda)$ grid for the $8^2$ lattice for three parameter groups; $\theta_{\mathrm{zero\_mode}}$, $\theta_{\mathrm{scale}}$ and $\theta_{\mathrm{conv}}$. Overlaid are linear fits of the $\chi$ (green, dashed) and $\xi$ (blue, dotted) maxima, serving as transition proxies. The parameter groups ($\theta_{\mathrm{zero\_mode}}$, $\theta_{\mathrm{scale}}$) exhibit a sharp $A$-ridge tracking the critical line.}
\label{fig:phase_transition}
\end{figure*}

To identify where this deformation originates within the architecture, we partition the weights $\theta$ into six functional
groups (see \Cref{app:dg:cond}): two that act non-locally on
configurations (the zero mode scaling $\theta_{\mathrm{zero\_mode}}$
and the Fourier scaling $\theta_{\mathrm{scale}}$) and four bulk
groups ($\theta_{\mathrm{conv}}$, $\theta_{\mathrm{feat\_super}}$,
$\theta_{\mathrm{time\_kernel}}$, $\theta_{\mathrm{feat\_map}}$)
that act mostly locally.
Computing the area element per group,
$A_c(g) = \sqrt{\det G_c(g)}$, we find that only the non-local groups
develop a sharp ridge tracking the critical line, while the bulk
groups remain featureless (\Cref{fig:phase_transition}).
\Cref{tab:phase_correlations} confirms this quantitatively:
$\theta_{\mathrm{zero\_mode}}$ and $\theta_{\mathrm{scale}}$ exhibit
positive correlations between $A_c(g)$ and all three criticality
observables $\xi$, $|\nabla U_4|$, and $\chi$. The bulk groups, by
contrast, show no positive correlation with any criticality
observable; their dominant signal is a strong negative correlation
with $\lambda$ ($r \in (-0.76, -0.59)$), indicating that they track
the quartic coupling rather than proximity to the transition.

The physics is consistent with this observation: a transition
driven by the divergence of the correlation length must be absorbed
by the long-range degrees of freedom of the model. Restricted to those non-local groups, the geometry of the
parameter map recovers the phase boundary without drawing any
samples.

\begin{table}
\caption{Pearson correlation between the per-group area element
$A_c(g) = \sqrt{\det G_c(g)}$ and physical observables across the
$(m^2, \lambda)$ grid. Bold entries mark $r \geq 0.5$.
$\theta_\mathrm{bulk}$ aggregates the four bulk groups; the column
reports the range across these groups. Per-group values are given in
\Cref{tab:phase_correlations_full}.}
\label{tab:phase_correlations}
\centering
\small
\begin{tabular}{l c c c}
\toprule
Observable & $\theta_{\mathrm{zero\_mode}}$ & $\theta_{\mathrm{scale}}$ & $\theta_{\mathrm{bulk}}$ (range) \\
\midrule
$\xi$            & $\mathbf{+0.57}$ & $+0.36$          & $(-0.26, -0.12)$ \\
$|\nabla U_4|$   & $\mathbf{+0.61}$ & $\mathbf{+0.56}$ & $(-0.11, +0.07)$ \\
$\chi$           & $\mathbf{+0.67}$ & $\mathbf{+0.51}$ & $(-0.24, -0.06)$ \\
$|m|$            & $-0.04$          & $-0.10$          & $(-0.05, +0.14)$ \\
$\lambda$        & $-0.11$          & $-0.14$          & $(-0.76, -0.59)$ \\
\bottomrule
\end{tabular}
\end{table}

\begin{figure}
    \centering
    \begin{subfigure}[t]{0.48\linewidth}
        \subcaption{}
        \centering
        \includegraphics[width=\linewidth]{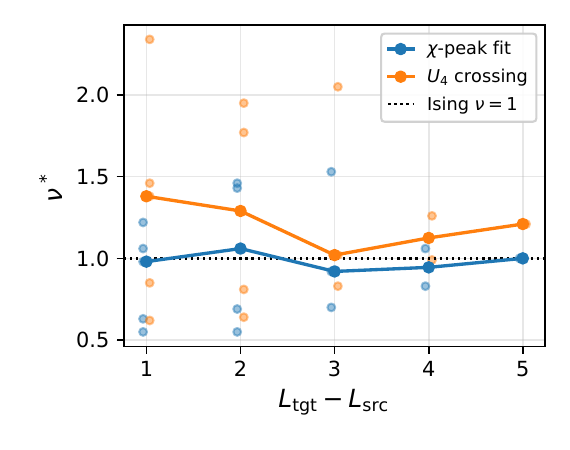}
        \label{fig:nu_trend}
    \end{subfigure}
    \hfill
    \begin{subfigure}[t]{0.48\linewidth}
        \subcaption{}
        \centering
        \includegraphics[width=\linewidth]{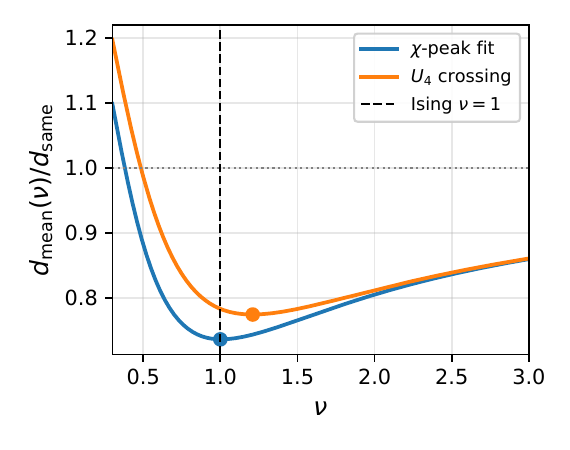}
        \label{fig:rg_6to11}
    \end{subfigure}
    \caption{{Finite-size-scaling exponent recovered from the latent geometry.
    \textbf{(a)} Seed-averaged minimizing exponent $\nu^*$ for the upscaling mappings on lattice sizes $L\in\{6,\dots,11\}$, as a function of the scale gap $L_\text{tgt}-L_\text{src}$. Points are individual pairings, the curves trace the per-gap medians for the two critical-line estimators of \Cref{app:critical_line}. The exponent is stable across scale gaps and brackets the exact Ising value $\nu=1$ (dotted line). 
    \textbf{(b)} Latent-distance curve for lattice sizes
    $6\to11$. Plots for more lattice pairs can be found in Appendix~\ref{app:fss}.}}
    \label{fig:rg_exponent}
\end{figure}

\subsection{Cross-lattice renormalization-group structure}
\label{sec:rg_latent}

The bare couplings $g=(m^2,\lambda)$ carry no absolute physical meaning: the same
$g$ describes different long-distance physics on lattices of different size $L$.
Relating theories across $L$ is precisely the content of the renormalization
group (RG), and near a continuous transition it is governed by finite-size
scaling (FSS). At fixed $\lambda$ and close to the critical line, the relevant
scaling field is
\begin{equation}
t = a(\lambda)\left(m^2 - m_c^2(\lambda)\right),
\label{eq:fss_variable}
\end{equation}
with $a(\lambda)$ a non-universal normalization, and FSS predicts that the
physics depends on $g$ and $L$ only through the dimensionless combination
$x = t\,L^{1/\nu}$, where $\nu$ is the correlation-length exponent. Two theories
at different $L$ sharing the same $x$ thus represent, up to finite-size
corrections, the same physics. For the two-dimensional $\phi^4$ theory studied
here $\nu=1$ (2D Ising universality), giving a ground-truth value against which
any learned notion of scale can be checked.

Requiring $x$ to be preserved between a source size $L_\text{src}$ and a target
size $L_\text{tgt}$, $t_\text{src}L_\text{src}^{1/\nu}=t_\text{tgt}L_\text{tgt}^{1/\nu}$,
and solving for the partner mass gives
\begin{equation}
m^2_{\mathrm{pred}} = m_c^2(\lambda;L_\text{tgt})
+ \left(m^2_\text{src}-m_c^2(\lambda;L_\text{src})\right)
\left(\frac{L_\text{src}}{L_\text{tgt}}\right)^{1/\nu}.
\label{eq:fss_map}
\end{equation}
Given a set of bare coupling constants on one lattice, \eqref{eq:fss_map}
predicts the bare couplings on the other lattice expected to share the same
physics.

\paragraph{Methodology.}
Our working hypothesis is that the JEPA encoder places physically similar
theories close together: if the latent geometry has captured the physics, weights
that are FSS partners across different lattice sizes should lie nearer in
embedding space than weights at unrelated couplings. For a pair
$(L_\text{src},L_\text{tgt})$ and a trial exponent $\nu$, \eqref{eq:fss_map}
assigns to each source coupling a partner coupling on the target lattice; we
measure the latent distance between the two, using the same-coupling pairing
($m^2(L_\text{src})=m^2(L_\text{tgt})$ held fixed) as a baseline. Because the FSS
partner itself depends on $\nu$, the form of the FSS (\ref{eq:fss_map}) allows us to read off the value of the critical exponent $\nu$ that describes the data the best. Evaluating \eqref{eq:fss_map} requires the critical line
$m_c^2(\lambda)$, which we estimate in two different ways---the susceptibility
($\chi$) peak and the Binder-cumulant crossing
(\Cref{app:critical_line}). These two definitions coincide in the continuum and
differ at finite $L$ only through their distinct numerical systematics, so that
the conclusion is robust against different sources of numerical error. For each
trial exponent we compute the mean latent distance between partners,
$d_{\mathrm{mean}}(\nu)=\tfrac1N\sum_i\lVert z_i^{(\text{src})}-z_i^{(\text{tgt})}\rVert_2$,
scan $\nu\in[0.2,3.5]$, and read off
$\nu^*=\arg\min_\nu d_{\mathrm{mean}}(\nu)$, with uncertainties from a bootstrap
over the $N{=}361$ couplings. Throughout, both lattices are embedded with the
encoder of the larger lattice, which is applied directly to the smaller-lattice
weight vectors without retraining (\Cref{app:rg_full}).

Two features of the setup limit the achievable precision, and we do not expect a
high-precision determination of $\nu$. The lattices are small, so finite-size
corrections to the leading-order relation \eqref{eq:fss_variable} are
appreciable; and locating the critical line $m_c^2(\lambda)$ is itself subject to
numerical error, which propagates into the predicted partner couplings.
Empirically these effects produce a sizable scatter in the per-pairing
minimizers, quantified in the Results below and in \Cref{app:rg_full}. We
therefore regard the analysis as a qualitative test of whether the latent
geometry is RG-aware, and not as a precise measurement of the critical exponent.

\paragraph{Results.}
In every pairing and for every critical-line estimator, FSS matching yields smaller latent distances than the same-coupling baseline (e.g.\ $d^*_{\mathrm{FSS}}/d_\text{same}\approx 0.72$ for $6\to11$); this confirms, as a basic consistency check, that the embedding responds to the scaling variable. More informatively, the minimizing exponent brackets the Ising value across the fifteen pairings on $L\in\{6,\dots,11\}$: with the $\chi$-peak critical line the per-pairing $\nu^*$ has an average of $0.98$, and the most widely separated pairing, $6\to11$, sits at $\nu^*=1.00_{-0.06}^{+0.08}$ (\Cref{fig:rg_6to11}). The individual estimates carry sizable uncertainty and depend on the critical-line definition---the averages over the two estimators via $\chi$-peak and $U_4$ crossing respectively span $0.55-1.53$ (median $0.98$) and $0.62-2.34$ (median $1.21)$---so we regard the outcome as broadly consistent with $\nu=1$ rather than a sharp measurement. The exponent is nonetheless stable across scale gaps (\Cref{fig:nu_trend}), and the latent-distance minima are clear and well separated from the same-coupling baseline (\Cref{fig:rg_all_scans}). Taken together, these results indicate that the
learned latent space captures the qualitative structure of the renormalization
group, organizing theories by their scaling variable across lattice sizes in a
way that is consistent with the exact $\nu=1$ of the 2D Ising class.
Full numerical results for all pairings and estimators are collected in \Cref{app:rg_full}.

As an ablation (Appendix~\ref{app:shuffled_control}) we perform a global label shuffle of the coupling grid, under which $(m^2, \lambda)$ are no longer decodable from the latent and the FSS minimum disappears. This provides further supporting evidence for the correctness of our methodology. 

\section{Conclusion}
We introduced JEPAWG, a JEPA-based weight generator that maps lattice couplings directly to flow parameters via a learned latent space, and
used it to study what neural networks learn about field theories from
their weights alone.
On 2D $\phi^4$ theory, the JEPAWG weight embedding recovers the intrinsic
dimensionality of the coupling manifold. Beyond this, the
geometry of the trained parameters exposes physical structure: the
pullback metric exhibits a ridge that traces the critical line of the
theory, attributed to the network parameters that are responsible for non-local
transformations on the generated configurations. Across lattice sizes, the
encoder organizes network weights according to the finite-size scaling variable predicted by the renormalization group. More quantitatively, the scaling exponent that yields the most coherent latent geometry is broadly consistent with the exact $\nu = 1$ prediction of the Ising model.
Notably,
none of these
properties are supervised by physical observables; they emerge from the
geometry of weight space and that of the JEPA embedding space
alone. As a generator, JEPAWG also produces functional flows at unseen couplings in a multi-seed setting where PCA, AE, and VAE baselines collapse.

From this work, the picture that emerges is a simple one: the geometry of trained weights reflects the geometry of the physical theory, and it concentrates on the components of the network where the physics actually lives. These results validate our JEPAWG framework and
support our broader claim on the effectiveness of physics as natural synthetic data for developing interpretable AI.

{\bf Limitations and Future Directions. } Our experiments are confined to a single 2D theory at small lattices, and our interpretability claims rest on agreement with one universality class. We see two natural extensions that we are currently pursuing. The first is
\emph{scale}: our experiments are confined to small lattices
($L \leq 11$) due to the cost of training the underlying flows, and
extending the analysis to larger lattices will further validate our results and in particular sharpen the weight-space RG results.
The second is \emph{diversity}: applying
JEPAWG to other quantum field theories, or other scientific systems would test whether the framework remains effective in richer physical settings. In favorable cases, the same approach could even uncover qualitative features of systems that are not yet fully understood theoretically, turning JEPAWG into an experimental tool.

\section*{Impact Statement}

This paper presents work whose goal is to advance the field of Machine Learning. There are many potential societal consequences of our work, none of which we feel must be specifically highlighted here.

\section*{Acknowledgments}
The work of Julian Ebelt, Tobias Göbel, Zier Mensch, Mathis Gerdes, and Miranda Cheng was supported by the AI-application grant AS-IAIA-114-M02 from Academia Sinica, and the Vici grant (number VI.C.232.117) from the Dutch Research Council (NWO).
Mathis Gerdes was also partially supported by the National Science Foundation under Cooperative Agreement PHY-2019786 (The NSF AI Institute for Artificial Intelligence and Fundamental Interactions, http://iaifi.org/).
Finally, we are grateful for the Wistron Computing Power Donation Program for compute support.
\bibliography{references}
\bibliographystyle{icml2026}

\newpage
\appendix
\onecolumn

\section{Experimental Details}
\subsection{Baseline Methods}
\label{app:baseline_methods}
\paragraph{Weight-space PCA.}
We perform PCA on the standardized training set $\{\tilde{\theta}_i\}_{i=1}^N$, retaining the top-$k$ principal components $U_k \in \mathbb{R}^{d_\theta \times k}$ and representing each parameter vector by its projection $\mathbf{z}_i = U_k^\top \tilde{\theta}_i$.

\paragraph{Autoencoder and variational autoencoder.}
\label{app:vae}
We train a symmetric MLP encoder $e_\phi : \mathbb{R}^{d_\theta} \to \mathbb{R}^{d_z}$ and decoder $d_\xi : \mathbb{R}^{d_z} \to \mathbb{R}^{d_\theta}$ with a Gaussian reparameterized latent $z = \mu_\phi(\tilde{\theta}) + \sigma_\phi(\tilde{\theta}) \odot \epsilon$, $\epsilon \sim \mathcal{N}(0, I)$, using the VAE objective~\citep{kingma2014auto}
\begin{equation}
  \mathcal{L} =
  \frac{1}{N}\sum_{i=1}^N
    \bigl\| d_\xi(z_i) - \tilde{\theta}_i \bigr\|_2^2
  + \beta\, D_{\mathrm{KL}}\!\bigl(
    q_\phi(z|\tilde{\theta}_i) \| \mathcal{N}(0,I)
  \bigr).
\end{equation}
Setting $\beta = 0$ gives the deterministic autoencoder (AE); $\beta = 1$ gives the standard VAE. In both cases, the embedding of point $i$ is the encoder mean $\mathbf{z}_i = \mu_\phi(\tilde{\theta}_i)$, and no coupling information is used during training.

\paragraph{Conditional generation.}
At inference, all three baselines introduce coupling information through a shared post-hoc step: we fit a 2-layer MLP from couplings to the learned weight embeddings. Given a new coupling $g^\star$, we predict $\hat{\mathbf{z}}(g^\star)$ and apply the corresponding inverse to recover parameters.

In contrast, JEPAWG requires no post-hoc fit: the coupling encoder $E_g$ and predictor $P$ are trained jointly with the weight encoder via the VICReg invariance, so the learned composition $P \circ E_g$ plays the role of the post-hoc regressor used by the baselines.

\subsection{VICReg objective}
\label{app:vicreg}

For a batch of $N$ paired samples, we denote by
$Z_x, Z_y \in \mathbb{R}^{N \times d_z}$ the stacked latent
representations produced by the two encoders, where each row
corresponds to one sample and each column to one latent dimension.
The predictor is applied row-wise per sample, yielding
$\hat{Z}_y = h(Z_x) \in \mathbb{R}^{N \times d_z}$.

The VICReg loss~\citep{bardes2022vicreg} is defined as
\begin{equation}
    \label{eq:vicreg}
    \mathcal{L} = \alpha\, s(Z_x, Z_y)
    + \beta\bigl[v(Z_x) + v(Z_y)\bigr]
    + \gamma\bigl[c(Z_x) + c(Z_y)\bigr],
\end{equation}
with weights $\alpha = \beta = 25$ and $\gamma = 1$.

\paragraph{Invariance term.}
The invariance term encourages alignment between predicted and target
representations,
\begin{equation}
    s(Z_x, Z_y)
    = \frac{1}{N} \sum_{i=1}^N
      \left\| \hat{\mathbf{z}}_{y,i} - \mathbf{z}_{y,i} \right\|_2^2,
\end{equation}
where $\hat{\mathbf{z}}_{y,i}$ and $\mathbf{z}_{y,i}$ denote the
$i$-th rows of $\hat{Z}_y$ and $Z_y$, respectively.

\paragraph{Variance term.}
Let $z^j \in \mathbb{R}^N$ denote the $j$-th column of $Z$, i.e.\ the
values of latent dimension $j$ across the batch. The variance term
penalizes dimensions whose empirical standard deviation falls below
one,
\begin{equation}
    v(Z)
    = \frac{1}{d_z} \sum_{j=1}^{d_z}
    \max\!\left(0,\; 1 - \sqrt{\operatorname{Var}(z^j) + \varepsilon}\right),
\end{equation}
where
\begin{equation}
    \operatorname{Var}(z^j)
    = \frac{1}{N} \sum_{i=1}^N
      \bigl(Z_{ij} - \mu_j\bigr)^2,
    \quad
    \mu_j = \frac{1}{N} \sum_{i=1}^N Z_{ij}.
\end{equation}
This term prevents representational collapse by ensuring that each
latent dimension varies across the batch.

\paragraph{Covariance term.}
Let $\bar{Z} \in \mathbb{R}^{N \times d_z}$ denote the mean-centered
representations,
\begin{equation}
    \bar{Z}_{ij} = Z_{ij} - \mu_j.
\end{equation}
The covariance matrix of the latent representation is then
\begin{equation}
    C(Z) = \frac{1}{N-1} \bar{Z}^\top \bar{Z} \in \mathbb{R}^{d_z \times d_z}.
\end{equation}
The covariance term penalizes correlations between distinct latent
dimensions,
\begin{equation}
    c(Z)
    = \frac{1}{d_z} \sum_{i \neq j} \bigl[C(Z)\bigr]_{ij}^2.
\end{equation}
This encourages different latent dimensions to capture independent
factors of variation.

\paragraph{Covariance ablation.}
Throughout the main text we set $\gamma = 0$, omitting the covariance
term so the latent space can concentrate onto fewer dimensions when
the data manifold admits a lower-dimensional structure
(\Cref{sec:jepawg}). For comparison, we also report results for the
standard VICReg setting $\gamma = 1$, denoted JEPAWG$^*$ in the tables. The
JEPAWG configuration thus refers to the covariance-off variant
combined with the decoder $D_\theta$.

\subsection{Weight symmetries and canonicalization}
\label{app:canon}
The factorization~\eqref{eq:factorization} introduces bond indices $d'$ and $f'$ that are contracted out, so the triple $(\widetilde{W}, W^K, W^H)$ is determined only up to invertible changes of basis: for any $M \in \mathrm{GL}(F')$, replacing $W^H \mapsto M\,W^H$ and contracting $\widetilde{W}$ with $(M^{-1})^\top$ on $f'$ leaves the physical weight tensor $W_{xydf}$ invariant, and analogously for $\mathrm{GL}(D')$ on the time bond. Two further symmetries act on the Fourier index $f \geq 2$: permutations $\sigma \in S_{F-1}$ of the pairs $(\omega_f, W^H_{\cdot,f})$, and paired sign flips $\omega_f \mapsto -\omega_f$, $W^H_{\cdot,f} \mapsto -W^H_{\cdot,f}$ (since $\sin$ is odd). Together these span a symmetry group
\begin{equation}
  G \;=\; \mathrm{GL}(F') \times \mathrm{GL}(D')
        \times S_{F-1} \times \{\pm1\}^{F-1},
\end{equation}
whose continuous part $\mathrm{GL}(F') \times \mathrm{GL}(D')$ has dimension $(F')^2 + (D')^2$, and under which distinct parameter vectors $\theta$ can encode the same input-output map.

We select a unique orbit representative as follows. Compute the QR decomposition $W^H = QR$ with $Q \in \mathbb{R}^{F' \times F'}$ orthogonal and $R \in \mathbb{R}^{F' \times F}$ upper trapezoidal, using the convention $\operatorname{diag}(R) \geq 0$ to fix the sign ambiguity. We partition $R = [\,R_1 \mid R_2\,]$ into its first $F'$ columns $R_1 \in \mathbb{R}^{F' \times F'}$ and the remaining columns $R_2 \in \mathbb{R}^{F' \times (F-F')}$, where $[\,\cdot \mid \cdot\,]$ denotes horizontal concatenation. Since $Q \in \mathrm{O}(F') \subset \mathrm{GL}(F')$, the map $W^H \mapsto Q^\top W^H = R$ is a valid symmetry transformation. Assuming $W^H$ has full row rank $F'$ (generic for trained weights), $R_1$ is upper-triangular with strictly positive diagonal and hence invertible, so a second transformation $R \mapsto R_1^{-1} R = [\,I_{F'} \mid R_1^{-1} R_2\,]$ brings $W^H$ to identity-block form. This selects the $\mathrm{GL}(F')$ orbit representative uniquely: if two matrices on the same orbit are both in this form, $[\,I \mid A\,] = M[\,I \mid A'\,] = [\,M \mid MA'\,]$, hence $M = I$ and $A = A'$. The combined transformation composes on the $\widetilde{W}$ side as $\widetilde{W} \mapsto \widetilde{W} \cdot R_1^\top Q^\top$ on $f'$, which preserves~\eqref{eq:factorization}. The same construction applied to $W^K$ selects the $\mathrm{GL}(D')$ orbit representative. Finally, the discrete factor $S_{F-1} \times \{\pm1\}^{F-1}$ is canonicalized by sorting $\{|\omega_f|\}_{f \geq 2}$ in ascending order and flipping signs to enforce $\omega_f \geq 0$, with columns of $W^H$ permuted and sign-flipped accordingly.

\section{Data Generation}
\label{app:data-generation}

The dataset analyzed in this work is a collection of \emph{trained network
parameters} of the coupling-conditional flow of \Cref{subsec:cond-flow}, each
near-optimal as a sampler for the two-dimensional $\phi^{4}$ lattice field
theory at a different point of the coupling space
$g=(m^{2},\lambda)$. We generate it in two stages: a
coupling-conditional pre-training that produces a single model covering the
whole coupling box (\Cref{app:dg:cond}), followed by per-point fine-tuning
on a regular grid. The output is an $N\times N$ array
of parameter vectors $\{\theta_{ij}\}_{i,j=1}^{N}$, all in the same
architecture and indexed by the underlying coupling $g_{ij}=(m^{2}_{i},\lambda_{j})$.

Throughout the paper we restrict the couplings to the box $\Omega=[-5.1,-1.9]\times[2.85,5.65]\ni(m^{2},\lambda)$, which straddles the symmetric and broken phases of the model. The single-lattice experiments (weight generation, phase structure, intrinsic dimension) use lattice sizes $L\in\{6,8,10\}$; the cross-lattice finite-size-scaling analysis of \Cref{sec:rg_latent} uses the full ladder $L\in\{6,\dots,11\}$. For each lattice size in the cross-lattice finite-size-scaling analysis, we train a separate coupling-conditional flow, all warm-started from a common base checkpoint and trained for the same number of steps, and then produce the grid by per-point fine-tuning.

\subsection{Coupling-conditional pre-training}
\label{app:dg:cond}
We build on the continuous normalizing flow architecture for sampling from the $\phi^4$ lattice field theory \citep{gerdes2023learning, Gerdes2025Bijx}. The flow is preceded by a Fourier-space preconditioner: a learnable per-mode diagonal rescaling in Fourier space, whose form is motivated by the free-theory propagator, which diagonalizes the quadratic action. Writing $\hat\phi(k)$ for the FFT of $\phi$, the preconditioner takes the form

\begin{equation}
    \hat\phi(k) \;\mapsto\; e^{s_k}\, \hat\phi(k),
\end{equation}
    with one learnable real scalar $s_k$ per reduced Fourier slot, while the zero mode $\hat\phi(0) \propto \sum_x \phi(x)$ is additionally mapped through a stack of $1$-D $\sinh$-based bijections \citep{gerdes2026analyticbijectionssmoothinterpretable} followed by a global scaling. This diagonal rescaling has exactly the form that reproduces the Gaussian per-mode variance $\frac{1}{2(\hat k^2 + m^2)}$ of the quadratic action; rather than fixing the $s_k$ to those free values, we let them be learned, so the preconditioner can settle at the effective per-mode variances of the full interacting theory.
    Unlike the neural-ODE factors, the preconditioner is not conditioned on $g$: a single set of scales $s_k$ and zero-mode parameters is shared across the whole coupling family in the pre-trained model. Its variation across the fine-tuned grid arises solely from the subsequent per-point fine-tuning.
    Note that the number of parameters in this layer depends on the lattice size $L$. We therefore align networks across lattice sizes to a common parameter shape by zero-padding smaller lattices and truncating larger ones in Fourier space.

The consecutive neural ODE then defines an invertible map $f_\theta \colon \mathbb{R}^{L^2} \to \mathbb{R}^{L^2}$ as the solution to a vector field of the form
\begin{equation}
  \frac{d\phi(t)_x}{dt}
  = \sum_{y,d,f} W_{xydf}\, K(t)_d\, H(\phi(t)_y)_f\,,
\end{equation}
where $H$ is a field basis of dimension $F$ defined by $H(\phi)_1 = \phi$ and $H(\phi)_f = \sin(\omega_f \phi)$ for $f \geq 2$, with learnable frequencies $\omega \in \mathbb{R}^{F-1}$, $K$ a Fourier time kernel of dimension $D$, and $W$ a learnable weight tensor constrained by the lattice symmetry group $C_L^2 \rtimes D_4$. To reduce the number of free parameters, $W$ is parameterized as
\begin{equation}
  W_{xydf} = \sum_{d',f'} \widetilde{W}_{xyd'f'}\,
  W^K_{d'd}\, W^H_{f'f}\,.
\label{eq:factorization}
\end{equation}

\paragraph{Parameter groups.}
The flow's parameter vector $\theta$ decomposes naturally into six
groups, referenced by name in \Cref{sec:phasetransition} and
\Cref{tab:phase_correlations}:
\begin{itemize}
\item $\theta_{\mathrm{zero\_mode}}$: parameters of the dedicated
  zero-mode flow acting on $\hat\phi(0)$, governing the
  global magnetization mode and its bimodal distribution in the
  broken phase.
\item $\theta_{\mathrm{scale}}$: the per-mode preconditioner scalars
  $s_k$ for $k \neq 0$, which set the overall variance of each
  Fourier mode.
\item $\theta_{\mathrm{conv}}$: the spatial weight tensor
  $\widetilde{W}_{xyd'f'}$ in~\eqref{eq:factorization}, which acts
  as a learnable convolution on the lattice.
\item $\theta_{\mathrm{time\_kernel}}$: the time-kernel factor
  $W^K_{d'd}$ in~\eqref{eq:factorization}.
\item $\theta_{\mathrm{feat\_super}}$: the feature-superposition
  factor $W^H_{f'f}$ in~\eqref{eq:factorization}.
\item $\theta_{\mathrm{feat\_map}}$: the learned frequencies
  $\omega \in \mathbb{R}^{F-1}$ that define the field basis $H$.
\end{itemize}
The first two groups parametrize global, non-local degrees of freedom
(zero mode and per-mode scale); the remaining four parametrize the
local, mode-mixing operator inside the neural ODE.

To learn a continuous family of theories parametrized by couplings $g = (m^2, \lambda)$, each factor in~\eqref{eq:factorization} is extended with an additional index contracted against an embedding $J(g) \in \mathbb{R}^A$, making the weight tensor $W_{xydf}(g)$ a smooth function of the theory parameters. At fixed couplings, the conditioned flow is fully determined by a parameter vector $\theta(g)$ obtained by flattening the factors $(\widetilde{W}, W^K, W^H)$ and the learned frequencies~$\omega$, together with the preconditioner parameters ($\theta_\mathrm{scale}$,
$\theta_\mathrm{zero\_mode}$). The JEPAWG weight encoder $E_\theta$ is applied
to the neural ODE factors $(\widetilde{W}, W^K, W^H, \omega)$ only
($d_\theta = 19{,}689$); the preconditioner is excluded from the embedding
because it is nearly invariant across fine-tuned couplings. We verified that including the preconditioner in the embedding leaves the encoder-based results essentially unchanged (FSS exponent consistent with $\nu=1$; latent intrinsic dimension PCA-$95\%$ $=2$, MLE $\approx2.0$).

\paragraph{Embedding.} The embedding $J(g)\in\mathbb{R}^{A}$ that conditions
the factors of~\eqref{eq:factorization} is implemented as sinusoidal
positional features applied independently to each coupling, concatenated, and
projected by a single (orthogonally initialized) linear layer. Concretely,
for $g_{a}\in\{m^{2},\lambda\}$ we form
$\mathrm{PE}_{a}(g_{a})=[\sin(\bm{\omega}_{a}g_{a}),
\cos(\bm{\omega}_{a}g_{a}),g_{a}]$, with frequencies $\bm{\omega}_{a}$
on a logarithmic ladder up to a maximum position $g_{a}^{\max}$, then set
$J(g)=W[\mathrm{PE}_{1}(m^{2})\,\|\,\mathrm{PE}_{2}(\lambda)]+\bm{b}$. We
use $A=16$, $40$ Fourier features per coupling, and
$g^{\max}=(2.1,\,4.0)$.

\paragraph{Objective.} At each gradient step, we draw $M$ couplings
$g^{(1)},\dots,g^{(M)}$ uniformly in $\Omega$, and for each, $B$
samples $\phi^{(b)}\sim q_{\theta(g)}$ from the conditional flow. The loss is the sample-mean reverse Kullback--Leibler divergence between flow and target,
\begin{equation}
    \mathcal{L}
    = \frac{1}{M}\!\sum_{m=1}^{M}\frac{1}{B}\!\sum_{b=1}^{B}\!
      \Bigl[\log q_{\theta(g^{(m)})}\!\bigl(\phi^{(m,b)}\bigr)
            + S_{g^{(m)}}\!\bigl(\phi^{(m,b)}\bigr)\Bigr].
\label{eq:dg:revkl}
\end{equation}
Gradients flow through both the per-factor base tensors and the embedding
parameters $W,\bm{b}$ via the conditioning of~\eqref{eq:factorization}.

\paragraph{Optimization.} We optimize with Adam ($\beta_{1}{=}0.85$,
$\beta_{2}{=}0.9$), initial learning rate $3{\times}10^{-4}$, decayed
exponentially by $\tfrac{1}{2}$ every $8000$ steps. We train for $30{,}000$
steps with $M=8$ couplings per step and $B=32$ samples per coupling (total
batch size $256$). The neural ODE is integrated with \texttt{diffrax} using
the \texttt{Tsit5} adaptive stepper ($\text{atol}=\text{rtol}=10^{-5}$).
Convergence is monitored on a fixed set of reference couplings via the
per-coupling effective sample size,
\begin{equation}
\label{eqn:ess}
\mathrm{ESS}(g)=(\sum_{b}w^{(b)})^{2}/(N\sum_{b}(w^{(b)})^{2})    
\end{equation}
with $w^{(b)}=p_{g}(\phi^{(b)})/q_{\theta(g)}(\phi^{(b)})$ and $N$ the number of samples.

\paragraph{Fine-tuning loop.} The unconditional flow is fine-tuned at fixed
$g_{ij}$ by minimizing the single-coupling reverse KL,
\begin{equation}
    \mathcal{L}_{ij}(\theta)
    = \frac{1}{B}\sum_{b=1}^{B}
      \Bigl[\log q_{\theta}(\phi^{(b)})+S_{g_{ij}}(\phi^{(b)})\Bigr],
    \quad \phi^{(b)}\sim q_{\theta}.
\end{equation}
We use Adam with initial learning rate $5{\times}10^{-4}$ (otherwise
identical betas), exponential decay with half-life $2000$ steps, batch size
$B=256$, and a step budget of $T_{\max}=1000$. Training stops as soon as a sufficient ESS is reached, or after $T_{\max}$ steps.

\paragraph{Compute.} All experiments ran on a single NVIDIA A100 or H100 GPU. Conditional flow pre-training takes a few hours; per-coupling fine-tuning, JEPAWG training, and the baselines each take on the order of minutes per run.

\subsection{Resulting dataset}
\label{app:dg:dataset}

The procedure above yields, for the box $\Omega=[-5.1,-1.9]\times[2.85,5.65]$ on lattices of size $6^2$ to $11^2$, a $19\times 19$
grid of fine-tuned flow parameter vectors, all in a common architecture and
all initialized from the same conditional model. By construction the
parameters vary smoothly across $\Omega$, a property inherited from the
conditional embedding and only mildly perturbed by the per-point
fine-tuning, which is the structural prerequisite for the analyses
presented in the main text.

\begin{figure}
    \centering
    \includegraphics[width=0.95\linewidth]{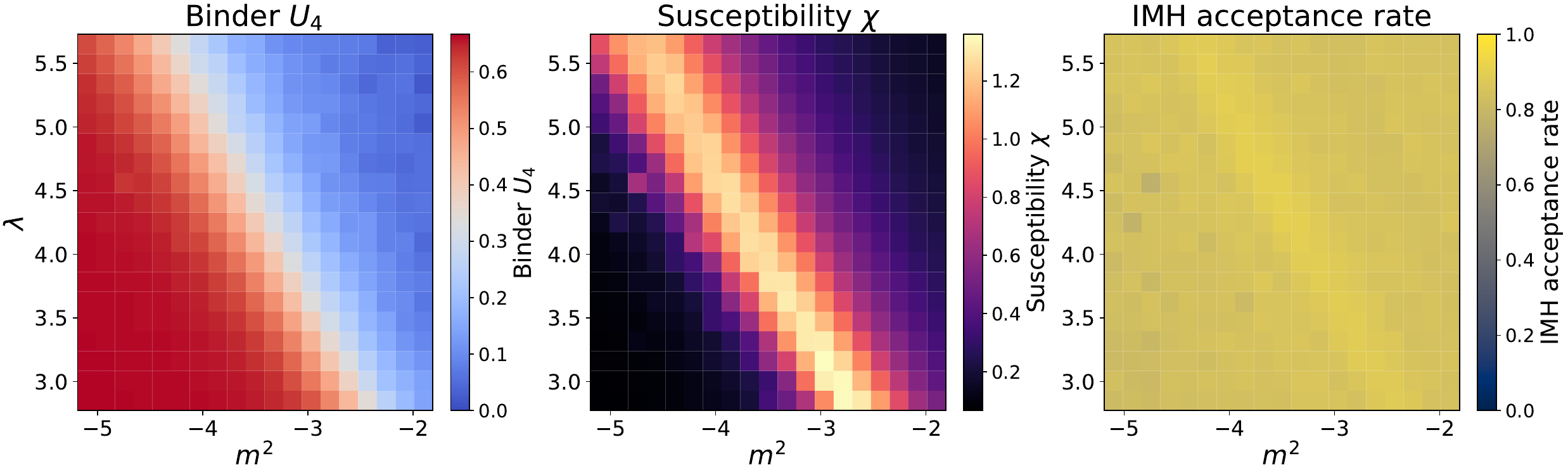}
\caption{Phase-structure observables on the $19{\times}19$ fine-tuned grid for the $8{\times}8$ lattice ($m^2 \in [-5.1,-1.9]$, $\lambda \in [2.85, 5.65]$). \textbf{Left:} Binder cumulant $U_4 = 1 - \langle\bar\phi^4\rangle/(3\langle\bar\phi^2\rangle^2)$. The high-$U_4$ ridge marks the broken phase, whereas the low-$U_4$ basin ($U_4\!\to\!0$) corresponds to the symmetric phase. The crossover between them traces out the critical line. \textbf{Center:} Susceptibility $\chi = L^2\!\left(\langle\bar\phi^2\rangle - \langle|\bar\phi|\rangle^2\right)$ peaks along the same crossover, as expected at a continuous phase transition. \textbf{Right:} Independent Metropolis Hastings acceptance rate at each grid point, computed from 4096 importance proposals from the fine-tuned conditional flow with a 256-step burn-in. Acceptance stays in the 0.76--0.91 range across the grid, confirming that the fine-tuned flow proposal remains a good match to the target distribution everywhere on the grid (no localized quality drops along the critical line). All quantities are estimated from samples drawn from the fine-tuned single-coupling flows, with further unbiasing using Independent Metropolis Hastings.}
    \label{fig:dg:obs}
\end{figure}

\Cref{fig:dg:obs} summarizes the physical content of the dataset:
the Binder cumulant $U_{4}$ tracks the symmetric-to-broken transition, and
the susceptibility $\chi$ peaks along the same crossover ridge.

\subsection{Data Analysis Tools}
\label{app:analysis-tools}

\paragraph{Intrinsic dimension estimation.}
We estimate the intrinsic dimension of each learned representation
using the \emph{Maximum Likelihood Estimator}
(MLE)~\citep{levina2004maximum}, which assumes the data lies near a
$d$-dimensional manifold and estimates $d$ from the rate at which the
number of neighbors grows with radius. For a point $x_i$ with
$K$-th nearest neighbor distance $R_K(x_i)$ and $j$-th nearest
neighbor distance $R_j(x_i)$, the local dimension estimate is
\begin{equation}
  \hat{d}_K(x_i) = \left(
    \frac{1}{K-1} \sum_{j=1}^{K-1} \log \frac{R_K(x_i)}{R_j(x_i)}
  \right)^{-1},
\end{equation}
and the global estimate is the average over all points. We use
$K=10$ throughout. We also considered the TwoNN
estimator~\citep{facco2017estimating} but found it unreliable on our
data, producing estimates that did not stabilize with subsample size;
we therefore report only MLE results.
The estimator is applied after projecting each representation onto
its top-5 PCA components, a standard denoising step for noisy
high-dimensional manifolds.

\section{Single Seed Experiment}
\label{sec:single_seed}
We repeat the weight generation experiment using flow parameters from
a single conditional flow, removing the seed-boundary discontinuity
present in the multi-seed setting of \Cref{sec:weight_generation}.
The training data consists of a $19 \times 19$ grid of fine-tuned
flows over the coupling box $\Omega$, all initialized from the same
conditional model. \Cref{tab:ess_single_seed} reports the decoded-flow
ESS across 9 regions of the coupling plane, averaged over 5 random
seeds.

In the single-seed setting, all methods benefit from the smooth
weight manifold. On the interior, JEPAWG and JEPAWG$^*$ both achieve ESS
above $0.97$, and even PCA with an MLP head reaches $0.95$. The
baselines that failed in the multi-seed setting (VAE: $0.74$, AE:
$0.78$) now produce functional flows, confirming that
their earlier failure was due to the seed discontinuity rather than
insufficient model capacity.

\begin{figure}
    \centering
    \includegraphics[width=\linewidth]{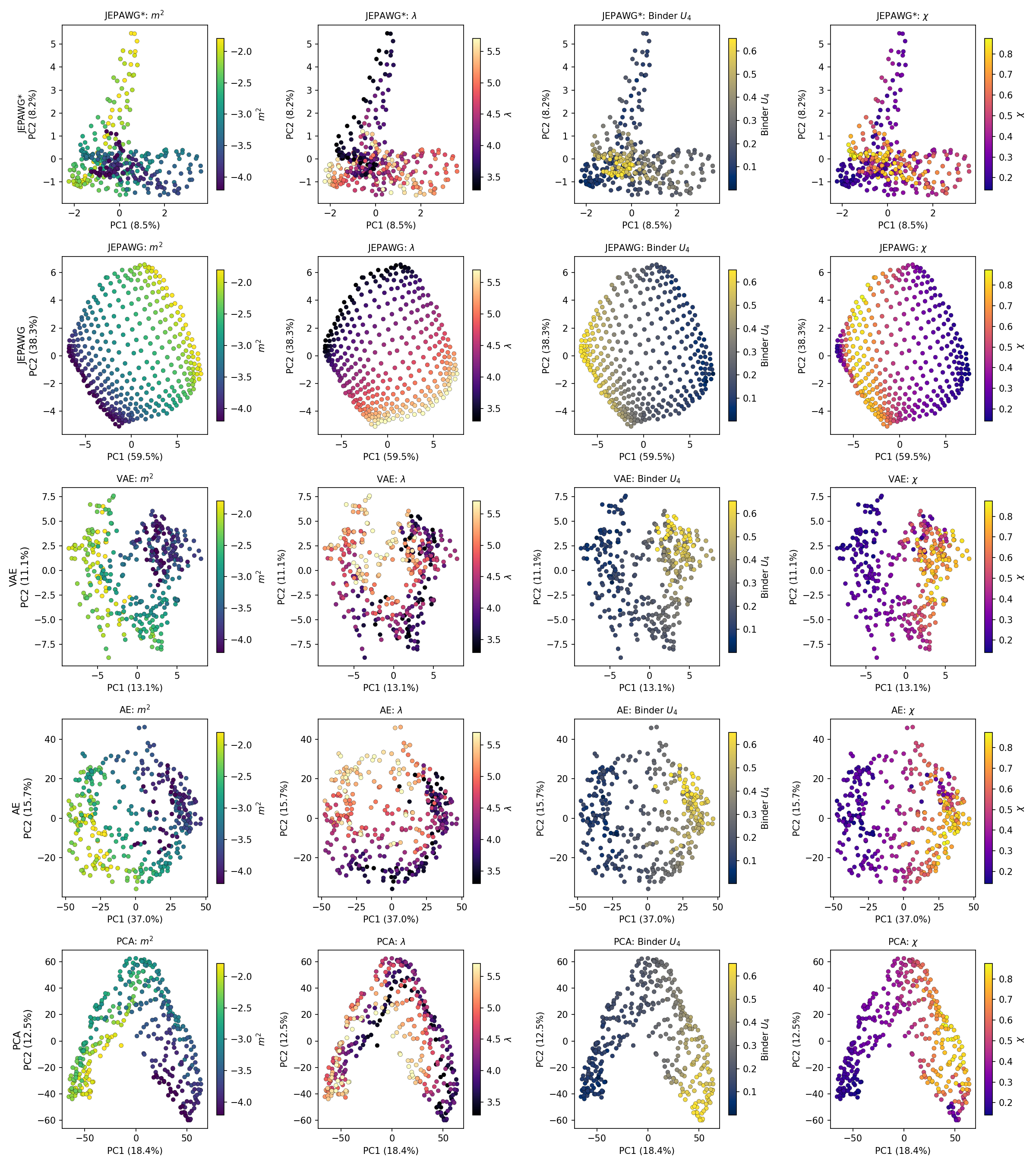}
    \caption{Latent-space embeddings (first two PCs) for JEPAWG$^*$, JEPAWG, VAE, AE, and PCA (top to bottom) in the single-seed setting, colored by $m^2$, $\lambda$, and the Binder cumulant $U_4$ (left to right).}
    \label{fig:single_seed_latents}
\end{figure}

\begin{figure}
    \centering
    \includegraphics[width=\linewidth]{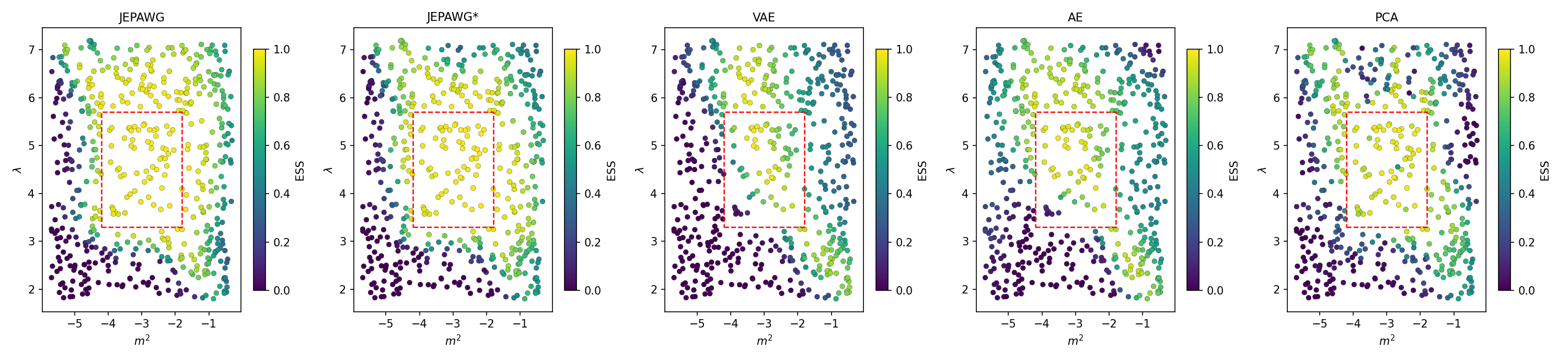}
    \caption{Per-coupling ESS scatter heatmaps for all methods in the single-seed
    setting, covering interpolation in the training region and
    extrapolation.}
    \label{fig:ess_single_seed_heatmap}
\end{figure}

On extrapolation, JEPAWG remains the strongest method overall
(\Cref{fig:ess_single_seed_heatmap}). Along the
$\lambda_{\mathrm{hi}}$ edge it reaches $0.92$, and at the
$(m^2_{\mathrm{hi}}, \lambda_{\mathrm{hi}})$ corner $0.73$, both
substantially above the baselines. Averaged over the four corner
regions, JEPAWG reaches an ESS of $0.51$, ahead of the strongest
baseline (VAE, $0.43$; \Cref{tab:ess_single_seed}). The
$(m^2_{\mathrm{lo}}, \lambda_{\mathrm{lo}})$ corner remains difficult
for all methods (ESS $\leq 0.07$). JEPAWG$^*$ performs comparably to JEPAWG
on the interior but falls behind on several extrapolation regions,
consistent with the observation that the covariance-off variant
concentrates its representation onto fewer latent dimensions and
extrapolates more smoothly. Embeddings for all methods are shown in
\Cref{fig:single_seed_latents}.
\begin{table}
\caption{Single-seed decoded-flow ESS across 9 regions of the
coupling plane. Interior: training window; edges and corners:
$\delta{=}1.5$ extrapolation. Each cell is the mean over 60
uniformly sampled evaluation couplings, $\pm$ standard deviation
across 5 random seeds. JEPAWG$^*$ refers to the covariance-on ablation
($\gamma = 1$); JEPAWG is the main configuration ($\gamma = 0$).
JEPAWG$^*$ and JEPAWG use the trained coupling predictor $P \circ E_g$; PCA, VAE, and AE use similarly sized MLP-head
$g \to \mathbf{z}$. VAE uses $\beta=1$, AE uses $\beta=0$.}
\label{tab:ess_single_seed}
\centering
\footnotesize
\setlength{\tabcolsep}{3pt}
\begin{tabular}{l ccccc}
\toprule
Region & PCA & JEPAWG$^*$ & JEPAWG & VAE & AE \\
\midrule
interior              & $0.947{\pm}0.007$ & $0.973{\pm}0.002$ & $\mathbf{0.976{\pm}0.001}$ & $0.737{\pm}0.030$ & $0.775{\pm}0.065$ \\
\midrule
$m^2_{\mathrm{lo}}$    & $0.287{\pm}0.056$ & $0.430{\pm}0.105$ & $\mathbf{0.440{\pm}0.046}$ & $0.148{\pm}0.070$ & $0.344{\pm}0.165$ \\
$m^2_{\mathrm{hi}}$    & $0.400{\pm}0.112$ & $0.667{\pm}0.134$ & $\mathbf{0.820{\pm}0.038}$ & $0.441{\pm}0.047$ & $0.395{\pm}0.201$ \\
$\lambda_{\mathrm{lo}}$ & $0.397{\pm}0.040$ & $0.386{\pm}0.056$ & $\mathbf{0.414{\pm}0.045}$ & $0.167{\pm}0.043$ & $0.174{\pm}0.021$ \\
$\lambda_{\mathrm{hi}}$ & $0.687{\pm}0.074$ & $0.825{\pm}0.076$ & $\mathbf{0.920{\pm}0.012}$ & $0.763{\pm}0.066$ & $0.781{\pm}0.041$ \\
\midrule
$(m^2_{\mathrm{lo}}, \lambda_{\mathrm{lo}})$ & $0.042{\pm}0.031$ & $\mathbf{0.065{\pm}0.045}$ & $0.059{\pm}0.016$ & $0.001{\pm}0.001$ & $0.029{\pm}0.028$ \\
$(m^2_{\mathrm{lo}}, \lambda_{\mathrm{hi}})$ & $0.311{\pm}0.154$ & $0.325{\pm}0.114$ & $0.599{\pm}0.079$ & $\mathbf{0.652{\pm}0.119}$ & $0.517{\pm}0.111$ \\
$(m^2_{\mathrm{hi}}, \lambda_{\mathrm{lo}})$ & $0.437{\pm}0.155$ & $0.412{\pm}0.188$ & $0.650{\pm}0.117$ & $\mathbf{0.738{\pm}0.078}$ & $0.498{\pm}0.190$ \\
$(m^2_{\mathrm{hi}}, \lambda_{\mathrm{hi}})$ & $0.239{\pm}0.150$ & $0.458{\pm}0.164$ & $\mathbf{0.732{\pm}0.036}$ & $0.329{\pm}0.066$ & $0.280{\pm}0.124$ \\
\midrule
corner avg.           & $0.257$ & $0.315$ & $\mathbf{0.510}$ & $0.430$ & $0.331$ \\
\bottomrule
\end{tabular}
\end{table}

\section{Finite-size scaling details}
\label{app:fss}

This appendix provides additional details for the finite-size scaling (FSS)
analysis discussed in \Cref{sec:rg_latent}, including the derivation of
the lattice-size matching rule, critical-line estimation procedures, and latent-distance scans for upscaling mappings.

\subsection{Derivation of the FSS partner map}
\label{app:fss_derivation}

Near a continuous phase transition, finite-size scaling predicts that
observables depend primarily on the scaling combination
\begin{equation}
x = t L^{1/\nu},
\end{equation}
where $t$ is the thermal scaling field and $\nu$ is the correlation-length
critical exponent. Close to the critical line, the thermal direction is
approximately aligned with the bare mass scaling field,
\begin{equation}
t \approx
a(\lambda)\left(m^2 - m_c^2(\lambda)\right),
\end{equation}
up to higher-order corrections and nonlinear mixing between scaling fields.

Theories with approximately equal values of the scaling variable
are expected to exhibit similar long-distance physics.
Therefore, to construct FSS partner points between a source lattice $L_\text{src}$ and a target lattice $L_\text{tgt}$, we require the scaling variable to be preserved,
\begin{equation}
t_\text{src}\,L_\text{src}^{1/\nu} = t_\text{tgt}\,L_\text{tgt}^{1/\nu}.
\label{eq:fss_appendix_constraint}
\end{equation}
Substituting the leading-order relation for the thermal scaling field and solving for the predicted partner mass on the target lattice yields
\begin{equation}
m^2_{\mathrm{pred}} = m_c^2(\lambda;L_\text{tgt}) + \left(m^2_\text{src} - m_c^2(\lambda;L_\text{src})\right)\left(\frac{L_\text{src}}{L_\text{tgt}}\right)^{1/\nu}.
\end{equation}
This construction defines the FSS partner point on the target lattice used to test whether latent-space proximity aligns with finite-size scaling trajectories.

\subsection{Critical-line estimation}
\label{app:critical_line}

The FSS matching procedure requires an estimate of the critical line $m_c^2(\lambda)$. We consider two estimators:

\begin{enumerate}
    \item \textbf{$\chi$-peak fit:} 
    for each $\lambda$ we locate the maximum of the magnetic susceptibility $\chi$ along the $m^2$ direction (with parabolic refinement) to obtain a pseudo-critical mass, average over lattice sizes, and fit the result with a low-order polynomial $m_c^2(\lambda)$.

    \item \textbf{$U_4$ crossing estimator:}
    the critical mass is obtained from Binder-cumulant crossings between neighboring lattice sizes.
    
\end{enumerate}

These estimators differ in their finite-size corrections and therefore provide a robustness check for the latent-space FSS analysis.

\subsection{Complete upscaling latent-distance scans}
\label{app:rg_full}

For each upscaling pairing, we scan the scaling exponent
$\nu \in [0.2,3.5]$ and compute the mean latent-space distance
\begin{equation}
d_{\mathrm{mean}}(\nu)
=
\frac{1}{N}
\sum_{i=1}^{N}
\left\|
z_i^{(\mathrm{src})}
-
z_i^{(\mathrm{tgt})}
\right\|_2.
\end{equation}

The optimal scaling exponent is defined by
\begin{equation}
\nu^*
=
\arg\min_\nu d_{\mathrm{mean}}(\nu),
\end{equation}
while
\begin{equation}
d^*_{\mathrm{FSS}}
=
\min_\nu d_{\mathrm{mean}}(\nu).
\end{equation}

\begin{figure}
    \centering
    \includegraphics[width=0.85\linewidth]{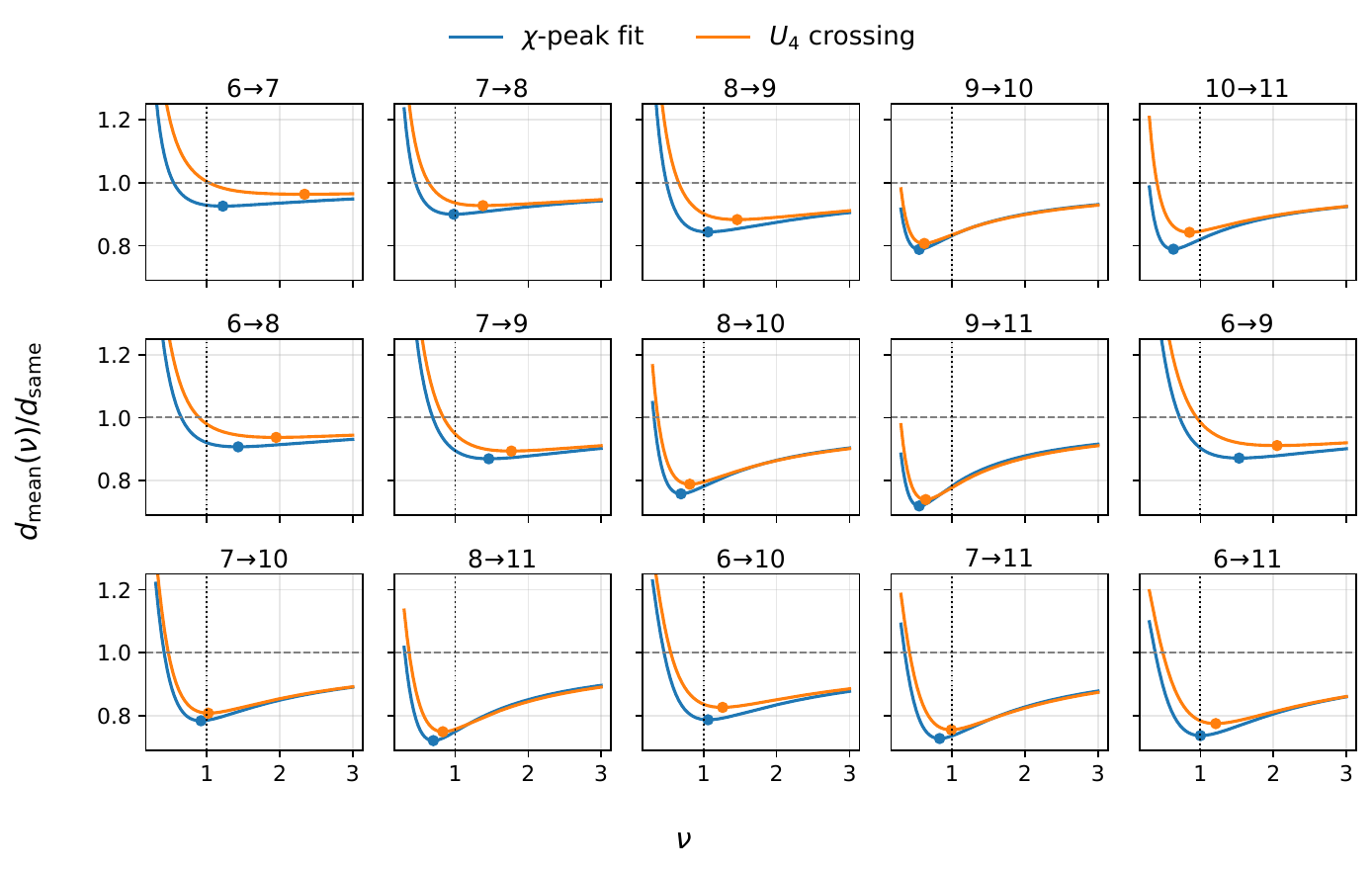}
    \caption{
        Latent-distance curves for all fifteen upscaling pairings ($L_\text{src}\to L_\text{tgt}$) on the $L\in\{6,\dots,11\}$ ladder. Each panel shows the mean latent distance $d_{\mathrm{mean}}(\nu)$ normalized by the same-coupling baseline, seed-averaged over the retained encoders, for the two critical-line estimators of \Cref{app:critical_line}. Filled circles mark the minimizer $\nu^*$; the dashed line is the same-coupling baseline and the dotted line marks the Ising value $\nu=1$.
    }
    \label{fig:rg_all_scans}
\end{figure}

\Cref{fig:rg_all_scans} shows the latent-distance curves for all upscaling pairings. \Cref{tab:rg_full} summarizes the corresponding numerical results. 

For each pairing the embeddings of both lattices are computed with the encoder of the larger lattice, applied to the smaller-lattice weights without retraining. 
 Partner couplings predicted outside the training window are discarded, with the same-coupling baseline recomputed on the identical retained set\footnote{We notice that JEPA training occasionally results in a collapsed latent space representing roughly only a single coupling direction. We discard these pathological training runs. An analysis of this seed dependence is delegated to future work. }. 

\paragraph{Uncertainty on $\nu^*$.}
We estimate the uncertainty on $\nu^*$ with a nonparametric bootstrap over the $N$ coupling points ($B{=}500$ resamples), taking the $[2.5,97.5]$ percentile interval of the resampled minimizers $\nu^*_b$. An interval away from the scan boundaries $\nu\in[0.2,3.5]$ indicates an interior minimum. The bootstrap captures only the sensitivity to the coupling points, not the systematic difference between the two critical-line estimators, which is visible from the two curves in each panel.

\begin{table}
\centering
\caption{
Finite-size-scaling results for all fifteen upscaling pairings on the $L\in\{6,\dots,11\}$ ladder, grouped by scale gap. Each entry is the seed-averaged minimizing exponent $\nu^*$ with its $95\%$ bootstrap confidence interval ($B{=}500$ resamples over the $N{=}361$ coupling points), for the two critical-line estimators of \Cref{app:critical_line}. The final row reports the median over all fifteen pairings.}
\label{tab:rg_full}
\begin{tabular}{ll cc}
\toprule
pair & gap & $\chi$-peak fit & $U_4$ crossing\\
\midrule
$6\!\to\!7$ & 1 & $1.22\,[0.98,1.56]$ & $2.34\,[1.75,3.00]$ \\
$7\!\to\!8$ & 1 & $0.98\,[0.88,1.09]$ & $1.38\,[1.19,1.56]$ \\
$8\!\to\!9$ & 1 & $1.06\,[0.94,1.21]$ & $1.46\,[1.26,1.69]$ \\
$9\!\to\!10$ & 1 & $0.55\,[0.51,0.60]$ & $0.62\,[0.56,0.68]$ \\
$10\!\to\!11$ & 1 & $0.63\,[0.59,0.67]$ & $0.85\,[0.78,0.95]$ \\
$6\!\to\!8$ & 2 & $1.43\,[1.20,1.65]$ & $1.95\,[1.65,2.31]$ \\
$7\!\to\!9$ & 2 & $1.46\,[1.33,1.64]$ & $1.77\,[1.55,2.01]$ \\
$8\!\to\!10$ & 2 & $0.69\,[0.64,0.74]$ & $0.81\,[0.75,0.89]$ \\
$9\!\to\!11$ & 2 & $0.55\,[0.52,0.58]$ & $0.64\,[0.59,0.68]$ \\
$6\!\to\!9$ & 3 & $1.53\,[1.33,1.76]$ & $2.05\,[1.71,2.45]$ \\
$7\!\to\!10$ & 3 & $0.92\,[0.85,0.98]$ & $1.02\,[0.94,1.12]$ \\
$8\!\to\!11$ & 3 & $0.70\,[0.67,0.74]$ & $0.83\,[0.78,0.90]$ \\
$6\!\to\!10$ & 4 & $1.06\,[0.98,1.16]$ & $1.26\,[1.13,1.43]$ \\
$7\!\to\!11$ & 4 & $0.83\,[0.78,0.88]$ & $0.99\,[0.92,1.05]$ \\
$6\!\to\!11$ & 5 & $1.00\,[0.94,1.08]$ & $1.21\,[1.12,1.31]$ \\
\midrule
\multicolumn{2}{l}{median over 15 pairs} & $0.98$ & $1.21$ \\
\bottomrule
\end{tabular}
\end{table}

\section{Phase Transition Detection}
\label{app:phase_transition}

\begin{figure}
\centering
\includegraphics[width=0.85\linewidth]{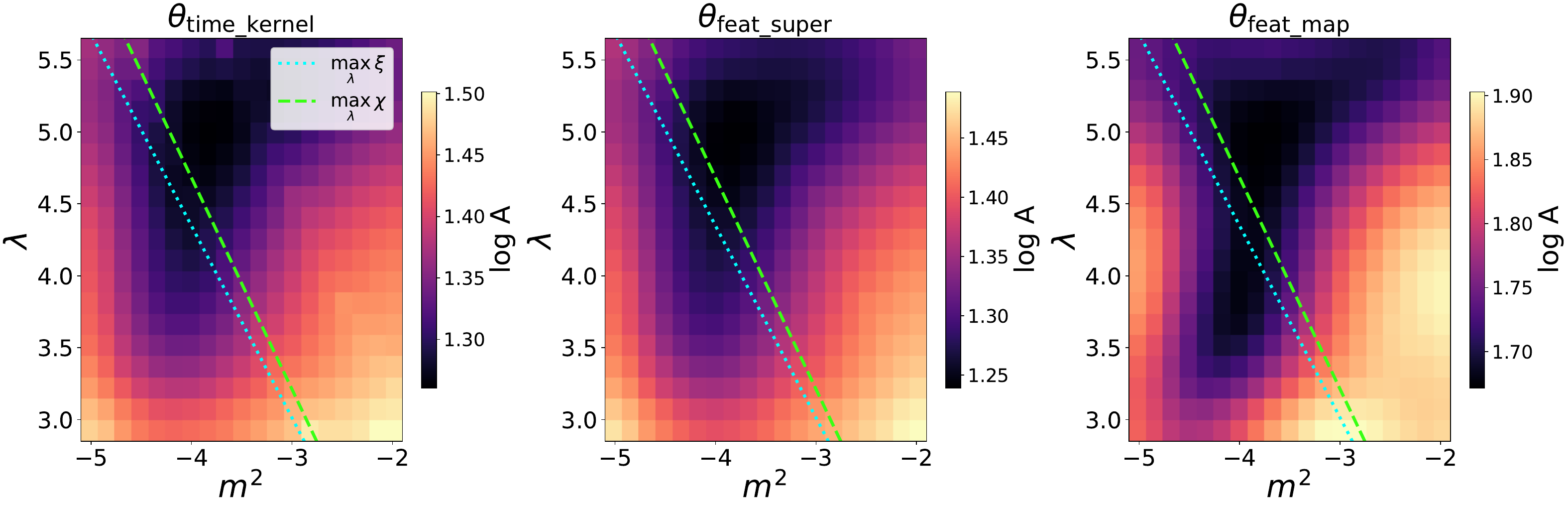}
\caption{Per-parameter-group pullback area $A(g)$ on the $(m^2, \lambda)$ grid for the $8^2$ lattice. Here, the remaining parameter-groups, complementary to the groups mentioned in \Cref{fig:phase_transition}. Overlaid are linear fits of the $\chi$ (green, dashed) and $\xi$ (blue, dotted) maxima, serving as transition proxies.}
\label{fig:phase_transition_bulk_components}
\end{figure}

\Cref{tab:phase_correlations_full} expands the aggregated bulk row of
\Cref{tab:phase_correlations} into per-group correlations between the
area element $A_c(g)$ and the physical observables. The four bulk
groups behave consistently: each shows a strong negative correlation
with $\lambda$ ($r \in [-0.76, -0.59]$) and no positive correlation
with any criticality observable. The non-local groups
$\theta_{\mathrm{zero\_mode}}$ and $\theta_{\mathrm{scale}}$ instead
correlate positively with $\xi$, $|\nabla U_4|$, and $\chi$, and
only weakly with $\lambda$. The complementary spatial maps for the
bulk groups are shown in
\Cref{fig:phase_transition_bulk_components}.

\begin{table}
\caption{Per-group Pearson correlations between the per-component
area element $A_c(g) = \sqrt{\det G_c(g)}$ and physical observables
across the $(m^2, \lambda)$ grid. Full breakdown of the four bulk
groups aggregated as $\theta_\mathrm{bulk}$ in \Cref{tab:phase_correlations}.
Bold entries mark $r \geq 0.5$.}
\label{tab:phase_correlations_full}
\centering
\small
\begin{tabular}{l c c c c c}
\toprule
 & $\xi$ & $|\nabla U_4|$ & $\chi$ & $|m|$ & $\lambda$ \\
\midrule
$\theta_{\mathrm{conv}}$         & $-0.12$ & $+0.07$ & $-0.06$ & $+0.11$ & $\mathbf{-0.75}$ \\
$\theta_{\mathrm{feat\_super}}$  & $-0.26$ & $-0.11$ & $-0.24$ & $+0.14$ & $\mathbf{-0.72}$ \\
$\theta_{\mathrm{time\_kernel}}$ & $-0.21$ & $-0.05$ & $-0.19$ & $+0.14$ & $\mathbf{-0.76}$ \\
$\theta_{\mathrm{feat\_map}}$    & $-0.20$ & $-0.03$ & $-0.17$ & $-0.05$ & $\mathbf{-0.59}$ \\
$\theta_{\mathrm{zero\_mode}}$   & $\mathbf{+0.57}$ & $\mathbf{+0.61}$ & $\mathbf{+0.67}$ & $-0.04$ & $-0.11$ \\
$\theta_{\mathrm{scale}}$        & $+0.36$          & $\mathbf{+0.56}$ & $\mathbf{+0.51}$ & $-0.10$ & $-0.14$ \\
\bottomrule
\end{tabular}
\end{table}

\section{Additional Data} 
\begin{figure}
    \centering
    \includegraphics[width=0.45\linewidth]{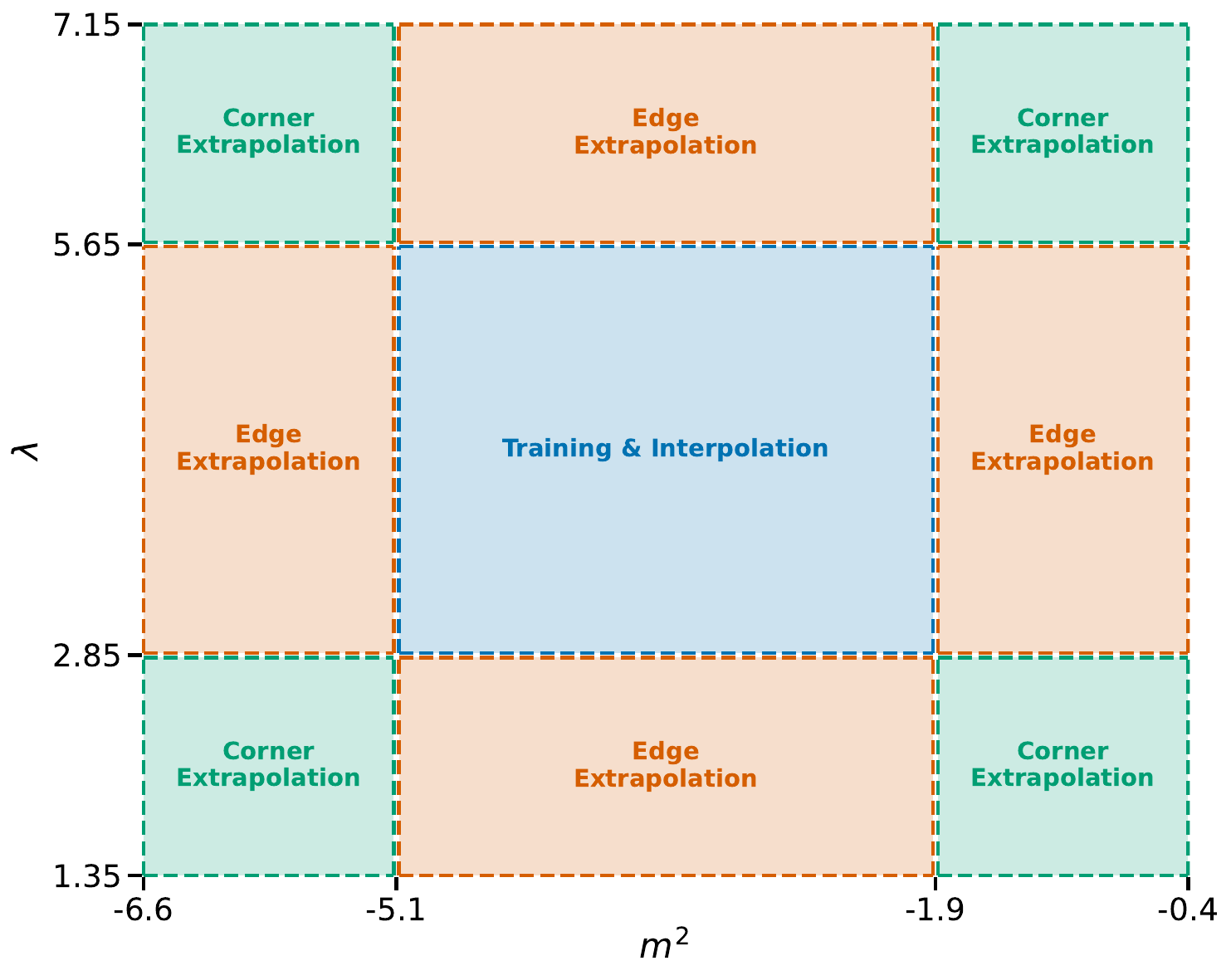}
    \caption{Split of the coupling space $(m^2,\lambda)$ into training and interpolation (blue), and extrapolation regions. The extrapolation regions are further divided into edges (orange) and corners (green).}
    \label{fig:box_explanation}
\end{figure}

\begin{figure}
\centering
\includegraphics[width=0.82\linewidth]{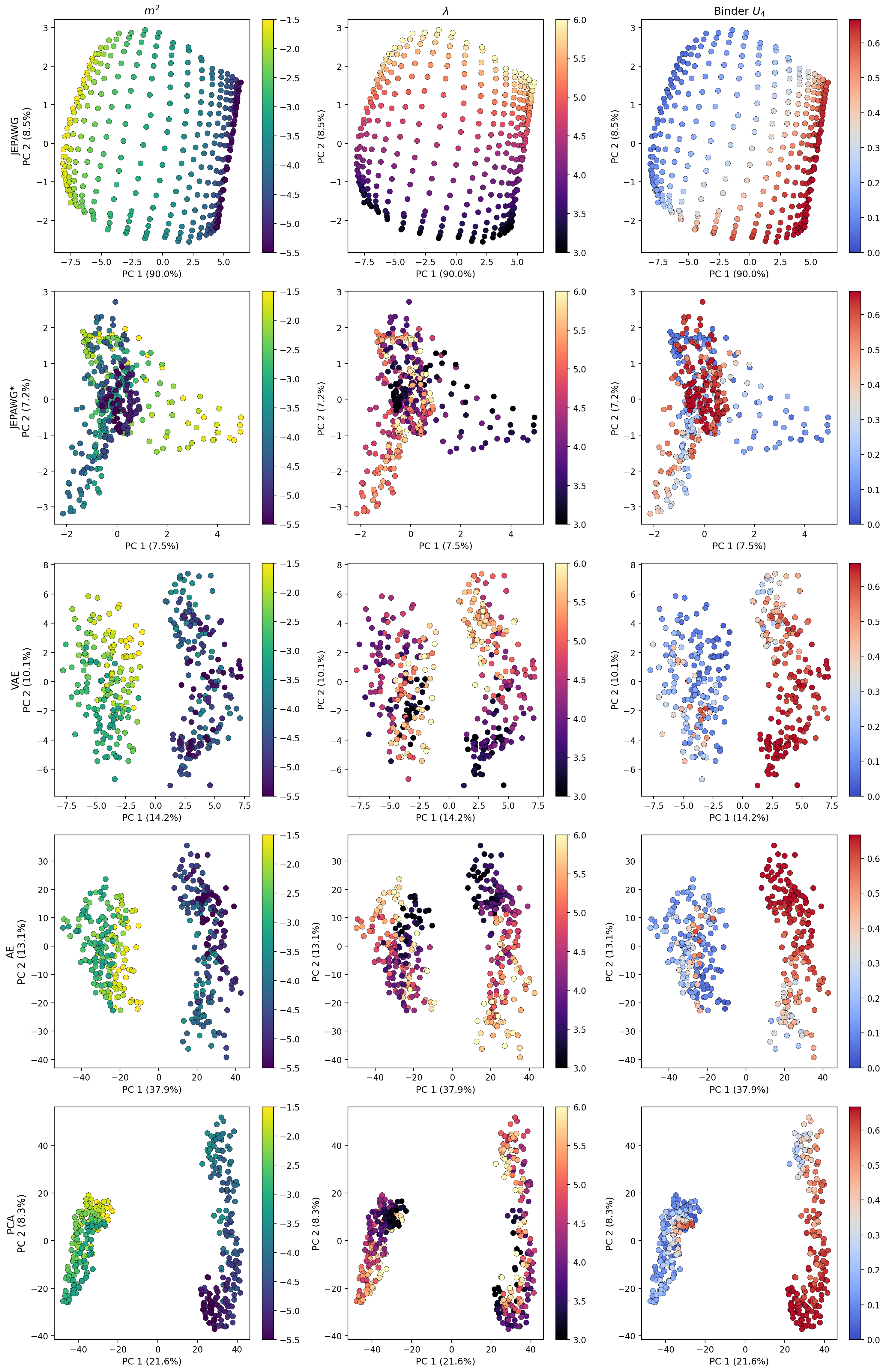}
\caption{PCA of latent representations ($d_z{=}16$) for JEPAWG, JEPAWG$^*$,
VAE, AE, and PCA (top to bottom), trained on 361 canonicalized flow
grid points. Points are colored by $m^2$, $\lambda$, and the Binder
cumulant $U_4$. JEPAWG organizes the latent space along axes aligned
with the coupling parameters and phase structure; the other methods
do not.}
\label{fig:all-methods-latent}
\end{figure}

\Cref{fig:box_explanation} visualizes the split of the coupling space $(m^2,\lambda)$ into training, interpolation, and extrapolation regions. \Cref{tab:ess_split_seed} reports decoded-flow ESS broken down across all 9 regions of the coupling plane, comparing canonicalized and raw weights for all five methods. \Cref{fig:all-methods-latent} extends the latent-space comparison of the main text to all five methods.

\begin{table}
\caption{Decoded-flow ESS on the split-seed grid across 9 regions of the coupling plane (6$\times$6, $m^2 \in [-5.1,-1.9]$, $\lambda \in [2.85,5.65]$). Interpolation: 256 random $(m^2,\lambda)$ over the joint window; edges and corners: $\delta{=}1.5$ extrapolation, 32 random points per region. Each cell is mean$\pm$std over the points in that region (single seed). JEPAWG$^*$ refers to the covariance-on ablation ($\gamma{=}1$); JEPAWG is the main configuration ($\gamma{=}0$). JEPAWG$^*$ and JEPAWG use the trained coupling predictor $P\circ E_g$; VAE, AE, and PCA use a MLP head $g \to \mathbf{z}$. VAE uses $\beta{=}1$, AE uses $\beta{=}0$. All evaluations use 4096 Metropolis Hastings samples per point.}
\label{tab:ess_split_seed}
\centering
\footnotesize
\setlength{\tabcolsep}{3pt}
\begin{tabular}{l ccccc}
\toprule
Region & PCA & JEPAWG$^*$ & JEPAWG & VAE & AE \\
\midrule
Can. Interp.                                 & $0.298{\pm}0.31$ & $0.613{\pm}0.32$ & $\mathbf{0.876{\pm}0.18}$ & $0.395{\pm}0.34$ & $0.327{\pm}0.33$ \\
\midrule
$m^2_{\mathrm{lo}}$                          & $0.127{\pm}0.22$ & $0.518{\pm}0.30$ & $\mathbf{0.881{\pm}0.06}$ & $0.325{\pm}0.29$ & $0.132{\pm}0.19$ \\
$m^2_{\mathrm{hi}}$                          & $0.091{\pm}0.19$ & $0.205{\pm}0.25$ & $\mathbf{0.439{\pm}0.37}$ & $0.133{\pm}0.27$ & $0.070{\pm}0.18$ \\
$\lambda_{\mathrm{lo}}$                      & $0.070{\pm}0.18$ & $0.303{\pm}0.37$ & $\mathbf{0.380{\pm}0.40}$ & $0.248{\pm}0.35$ & $0.188{\pm}0.29$ \\
$\lambda_{\mathrm{hi}}$                      & $0.071{\pm}0.16$ & $0.379{\pm}0.28$ & $\mathbf{0.532{\pm}0.34}$ & $0.208{\pm}0.29$ & $0.198{\pm}0.27$ \\
\midrule
$(m^2_{\mathrm{lo}}, \lambda_{\mathrm{lo}})$ & $0.048{\pm}0.10$ & $0.206{\pm}0.26$ & $\mathbf{0.216{\pm}0.32}$ & $0.072{\pm}0.21$ & $0.002{\pm}0.00$ \\
$(m^2_{\mathrm{lo}}, \lambda_{\mathrm{hi}})$ & $0.217{\pm}0.24$ & $0.278{\pm}0.32$ & $\mathbf{0.696{\pm}0.15}$ & $0.373{\pm}0.26$ & $0.120{\pm}0.24$ \\
$(m^2_{\mathrm{hi}}, \lambda_{\mathrm{lo}})$ & $0.082{\pm}0.17$ & $\mathbf{0.368{\pm}0.26}$ & $0.254{\pm}0.32$ & $0.035{\pm}0.12$ & $0.183{\pm}0.32$ \\
$(m^2_{\mathrm{hi}}, \lambda_{\mathrm{hi}})$ & $0.048{\pm}0.16$ & $0.138{\pm}0.27$ & $\mathbf{0.405{\pm}0.34}$ & $0.021{\pm}0.10$ & $0.019{\pm}0.09$ \\
\midrule
Raw Interp.                                  & $0.749{\pm}0.27$ & $0.409{\pm}0.32$ & $0.783{\pm}0.22$ & $0.910{\pm}0.18$ & $\mathbf{0.915{\pm}0.18}$ \\
\midrule
$m^2_{\mathrm{lo}}$                          & $0.398{\pm}0.28$ & $0.286{\pm}0.33$ & $0.828{\pm}0.12$ & $0.892{\pm}0.11$ & $\mathbf{0.930{\pm}0.06}$ \\
$m^2_{\mathrm{hi}}$                          & $0.126{\pm}0.28$ & $0.101{\pm}0.20$ & $0.318{\pm}0.35$ & $0.237{\pm}0.34$ & $\mathbf{0.384{\pm}0.32}$ \\
$\lambda_{\mathrm{lo}}$                      & $0.305{\pm}0.41$ & $0.152{\pm}0.19$ & $0.445{\pm}0.38$ & $\mathbf{0.480{\pm}0.42}$ & $0.477{\pm}0.41$ \\
$\lambda_{\mathrm{hi}}$                      & $0.155{\pm}0.30$ & $0.179{\pm}0.26$ & $0.536{\pm}0.22$ & $0.475{\pm}0.34$ & $\mathbf{0.668{\pm}0.29}$ \\
\midrule
$(m^2_{\mathrm{lo}}, \lambda_{\mathrm{lo}})$ & $0.041{\pm}0.10$ & $0.205{\pm}0.26$ & $0.208{\pm}0.31$ & $\mathbf{0.536{\pm}0.36}$ & $0.512{\pm}0.42$ \\
$(m^2_{\mathrm{lo}}, \lambda_{\mathrm{hi}})$ & $0.184{\pm}0.27$ & $0.119{\pm}0.16$ & $0.552{\pm}0.32$ & $\mathbf{0.691{\pm}0.26}$ & $0.663{\pm}0.26$ \\
$(m^2_{\mathrm{hi}}, \lambda_{\mathrm{lo}})$ & $0.120{\pm}0.28$ & $0.035{\pm}0.08$ & $\mathbf{0.231{\pm}0.33}$ & $0.221{\pm}0.32$ & $0.198{\pm}0.30$ \\
$(m^2_{\mathrm{hi}}, \lambda_{\mathrm{hi}})$ & $0.049{\pm}0.16$ & $0.105{\pm}0.21$ & $\mathbf{0.380{\pm}0.36}$ & $0.249{\pm}0.29$ & $0.323{\pm}0.28$ \\
\bottomrule
\end{tabular}
\end{table}

\section{Shuffled Data Control Experiment}
\label{app:shuffled_control}

To verify that JEPAWG recovers coupling structure from the flow weights rather than
exploiting architectural inductive biases, we train three shuffled-label controls.
In each, the 361 fine-tuned grid points are randomly permuted before training so that
the encoder sees a broken correspondence between couplings and weights.
\emph{Global shuffle}: a uniformly random permutation of all 361 points.
\emph{Local shuffle $k{=}5$ / $k{=}25$}: a greedy random matching in which each
coupling is swapped only with one of its $k$ nearest neighbors in normalized
$(m^2,\lambda)$ space, limiting how far the pairing deviates from the true layout.
The permutation is fixed at the start of training and not re-drawn during training.
Six independent shuffle seeds are run per condition; the model initialization seed
is held fixed across them.

All shuffled models reach the same training-time $R^2 \approx 0.97$ as the clean
model, confirming sufficient capacity.  To evaluate whether the learned latent space
encodes the true coupling, we apply each trained encoder to the \emph{original},
unshuffled grid points and measure $R^2$ of a linear coupling predictor
(Table~\ref{tab:shuffled_r2}). The global control collapses ($R^2 \approx 0.08$--$0.10$). The local controls retain substantial signal that decreases monotonically with neighborhood radius: $R^2 \approx 0.96$ for $k{=}5$ and $R^2 \approx 0.91$ for $k{=}25$, reflecting that very local swaps preserve much of the coupling--weight structure.

\begin{figure}
\centering
\includegraphics[width=0.4\linewidth]{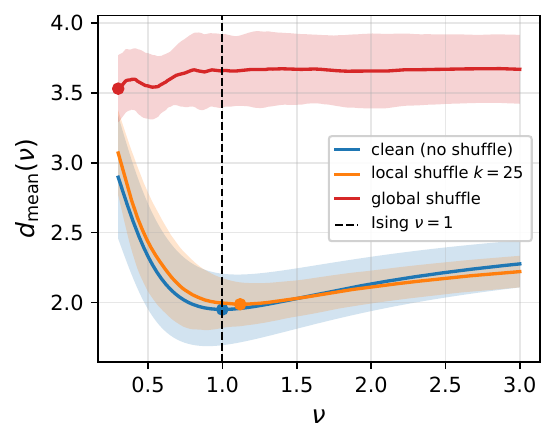}
\caption{
    FSS $\nu$ scan: mean latent distance $d(\nu)$ for the $6\to11$ lattice size pairing versus the trial exponent $\nu$, seed-averaged (shaded bands: $\pm1$ std over the six encoder seeds). The clean encoder and the local shuffle ($k{=}25$) reach low distances with a clear minimum near the Ising value. Under a global shuffle, which collapses the coupling decodability ($R^2\!\to\!0.09$, \Cref{tab:shuffled_r2}), the distances are roughly twice as large and stay high for all $\nu$: the embeddings no longer align, so no scaling exponent yields a minimum near $\nu=1$.
}
\label{fig:shuffled_nu}
\end{figure}

As a second check, we rerun the FSS $\nu$ scan on the shuffled embeddings (\Cref{fig:shuffled_nu}). When we shuffle all the couplings, the recovered exponent moves to the edge of the scan ($\nu^*\approx0.3$). When we only swap nearby couplings, the encoder still recovers the coupling and the exponent stays close to the unshuffled value.

\begin{table}
\caption{
  Shuffled-label control: $R^2$ of a linear coupling predictor applied to
  embeddings trained on shuffled grid points, then evaluated on the original
  (unshuffled) assignments.  Mean$\pm$std over 6 shuffle seeds; training-time
  $R^2$ is identical (${\approx}0.97$) for all conditions.
}
\label{tab:shuffled_r2}
\centering
\small
\begin{tabular}{lcc}
\toprule
Condition & $R^2(m^2)$ & $R^2(\lambda)$ \\
\midrule
Clean (no shuffle)      & $0.991$          & $0.988$          \\
Local shuffle $k{=}5$   & $0.973\pm0.003$  & $0.962\pm0.021$  \\
Local shuffle $k{=}25$  & $0.920\pm0.005$  & $0.919\pm0.004$  \\
Global shuffle          & $0.098\pm0.043$  & $0.081\pm0.041$  \\
\bottomrule
\end{tabular}
\end{table}

\end{document}